\documentstyle[sprocl,epsf]{article}

\def\spose#1{\hbox to 0pt{#1\hss}}
\def\lsim{\mathrel{\spose{\lower 3pt\hbox{$\mathchar"218$}}
 \raise 2.0pt\hbox{$\mathchar"13C$}}}
\def\gsim{\mathrel{\spose{\lower 3pt\hbox{$\mathchar"218$}}
 \raise 2.0pt\hbox{$\mathchar"13E$}}}

\begin{document}

\begin{titlepage}

\begin{flushright}
CERN-TH/97-24\\
hep-ph/9702375
\end{flushright}

\vspace{2cm}

\begin{center}
\Large\bf B Decays and the Heavy-Quark Expansion
\end{center}

\vspace{2cm}

\begin{center}
Matthias Neubert\\
{\sl Theory Division, CERN, CH-1211 Geneva 23, Switzerland}
\end{center}

\vspace{1cm}

\begin{abstract}
We review the theory and phenomenology of heavy-quark symmetry,
exclusive weak decays of $B$ mesons, inclusive decay rates and
lifetimes of $b$ hadrons.
\end{abstract}

\vspace{1.5cm}

\begin{center}
To appear in the Second Edition of\\
Heavy Flavours, edited by A.J. Buras and M. Lindner\\
(World Scientific, Singapore)
\end{center}

\vspace{2cm}

\noindent
CERN-TH/97-24\\
February 1997
\vfil

\end{titlepage}

\thispagestyle{empty}
\vbox{}
\newpage

\setcounter{page}{1}


\title{B DECAYS AND THE HEAVY-QUARK EXPANSION}

\author{MATTHIAS NEUBERT}

\address{Theory Division, CERN, CH-1211 Geneva 23, Switzerland}

\maketitle\abstracts{
We review the theory and phenomenology of heavy-quark symmetry,
exclusive weak decays of $B$ mesons, inclusive decay rates and
lifetimes of $b$ hadrons.}

\section{Introduction}

The rich phenomenology of weak decays has always been a source of
information about the nature of elementary particle interactions. A
long time ago, $\beta$- and $\mu$-decay experiments revealed the
structure of the effective flavour-changing interactions at low
momentum transfer. Today, weak decays of hadrons containing heavy
quarks are employed for tests of the Standard Model and measurements
of its parameters. In particular, they offer the most direct way to
determine the weak mixing angles, to test the unitarity of the
Cabibbo-Kobayashi-Maskawa (CKM) matrix, and to explore the physics
of CP violation. On the other hand, hadronic weak decays also serve
as a probe of that part of strong-interaction phenomenology which is
least understood: the confinement of quarks and gluons inside
hadrons.

The structure of weak interactions in the Standard Model is rather
simple. Flavour-changing decays are mediated by the coupling of the
charged current $J_{\rm CC}^\mu$ to the $W$-boson field:
\begin{equation}
   {\cal L}_{\rm CC} = - {g\over\sqrt{2}}\,J_{\rm CC}^\mu\,
   W_\mu^\dagger + \mbox{h.c.,}
\end{equation}
where
\begin{equation}
   J_{\rm CC}^\mu =
   (\bar\nu_e, \bar\nu_\mu, \bar\nu_\tau)\,\gamma^\mu
   \left( \begin{array}{c} e_{\rm L} \\ \mu_{\rm L} \\ \tau_{\rm L}
   \end{array} \right)
   + (\bar u_{\rm L}, \bar c_{\rm L}, \bar t_{\rm L})\,\gamma^\mu\,
   V_{\rm CKM} \left( \begin{array}{c} d_{\rm L} \\ s_{\rm L} \\
   b_{\rm L} \end{array} \right)
\end{equation}
contains the left-handed lepton and quark fields, and
\begin{equation}
   V_{\rm CKM} = \left( \begin{array}{ccc}
    V_{ud} & V_{us} & V_{ub} \\
    V_{cd} & V_{cs} & V_{cb} \\
    V_{td} & V_{ts} & V_{tb}
   \end{array} \right)
\end{equation}
is the CKM matrix. At low energies, the charged-current interaction
gives rise to local four-fermion couplings of the form
\begin{equation}\label{LFermi}
   {\cal L}_{\rm eff} = - 2\sqrt{2} G_F\,J_{\rm CC}^\mu
   J_{{\rm CC},\mu}^\dagger \,,
\end{equation}
where
\begin{equation}
   G_F = {g^2\over 4\sqrt{2} M_W^2} = 1.16639(2)~\mbox{GeV}^{-2}
\end{equation}
is the Fermi constant.

\begin{figure}[htb]
   \epsfxsize=6cm
   \centerline{\epsffile{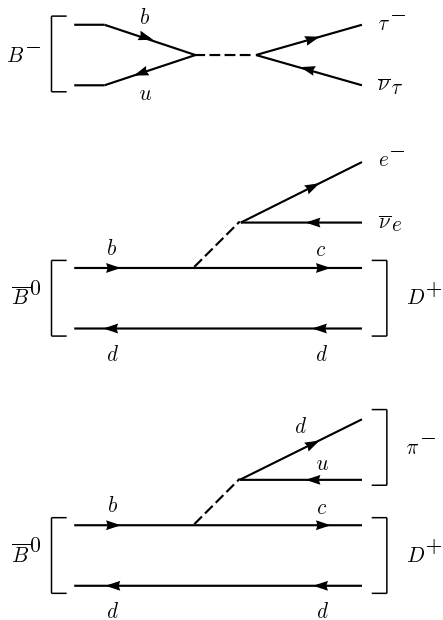}}
\caption{\label{fig:classes}
Examples of leptonic ($B^-\to\tau^-\bar\nu_\tau$), semileptonic
($\bar B^0\to D^+ e^-\bar\nu_e$), and non-leptonic ($\bar B^0\to
D^+\pi^-$) decays of $B$ mesons.}
\end{figure}

\begin{figure}[htb]
   \epsfxsize=8.5cm
   \centerline{\epsffile{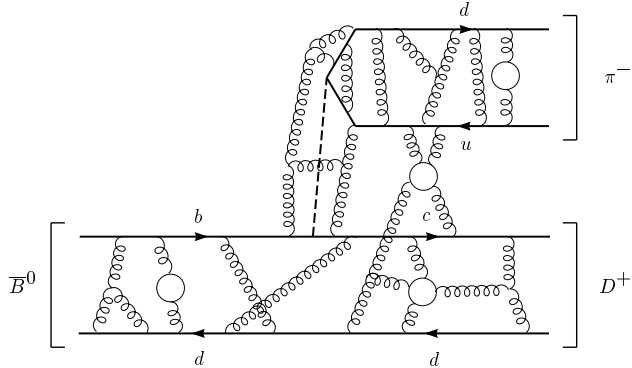}}
\caption{\label{fig:nonlep}
More realistic representation of a non-leptonic decay.}
\end{figure}

According to the structure of the charged-current interaction, weak
decays of had\-rons can be divided into three classes: leptonic
decays, in which the quarks of the decaying hadron annihilate each
other and only leptons appear in the final state; semileptonic
decays, in which both leptons and hadrons appear in the final state;
and non-leptonic decays, in which the final state consists of hadrons
only. Representative examples of these three types of decays are
shown in Fig.~\ref{fig:classes}. The simple quark-line graphs shown
in this figure are a gross oversimplification, however. In the real
world, quarks are confined inside hadrons, bound by the exchange of
soft gluons. The simplicity of the weak interactions is
over\-sha\-dowed by the complexity of the strong interactions. A
complicated interplay between the weak and strong forces
characterizes the phenomenology of hadronic weak decays. As an
example, a more realistic picture of a non-leptonic decay is shown in
Fig.~\ref{fig:nonlep}.

The complexity of strong-interaction effects increases with the
number of quarks appearing in the final state. Bound-state effects in
leptonic decays can be lumped into a single parameter (a ``decay
constant''), while those in semileptonic decays are described by
invariant form factors, depending on the momentum transfer $q^2$
between the hadrons. Approximate symmetries of the strong
interactions help us to constrain the properties of these form
factors. For non-leptonic decays, on the other hand, we are still far
from having a quantitative understanding of strong-interaction
effects even in the simplest decay modes.

Over the last decade, a lot of information on heavy-quark decays has
been collected in experiments at $e^+ e^-$ storage rings operating at
the $\Upsilon(4s)$ resonance, and more recently at high-energy $e^+
e^-$ and hadron colliders. This has led to a rather detailed
knowledge of the flavour sector of the Standard Model and many of the
parameters associated with it. There have been several great
discoveries in this field, such as $B^0$-$\bar B^0$
mixing~\cite{BBbar1,BBbar2}, $b\to u$
transitions~\cite{btou1}$^-$\cite{Bpirho}, and rare decays induced by
penguin operators~\cite{btos}. Yet there is much more to come. In
particular, the $B$-factories at SLAC, KEK, HERA-B and LHC-B will
provide a wealth of new results within the coming years.

The experimental progress in heavy-flavour physics has been
accompanied by a significant progress in theory, which was related to
the discovery of heavy-quark symmetry, the development of the
heavy-quark effective theory, and the establishment of the
heavy-quark expansion. The excitement about these developments rests
upon the fact that they allow (some) model-independent predictions in
an area in which ``progress'' in theory often meant nothing more than
the construction of a new model, which could be used to estimate some
strong-interaction hadronic matrix elements. In section~\ref{sec:2},
we explain the physical picture behind heavy-quark symmetry and
discuss the construction, as well as simple applications, of the
heavy-quark effective theory. Section~\ref{sec:3} deals with
applications of these concepts to exclusive weak decays of $B$
mesons. Applications of the heavy-quark expansion to the description
of inclusive decay rates and lifetimes of $b$ hadrons are the topics
of section~\ref{sec:4}.

\section{Heavy-Quark Symmetry}
\label{sec:2}

This section provides an introduction to the ideas of heavy-quark
symmetry~\cite{Shu1}$^-$\cite{Isgu} and the heavy-quark effective
theory~\cite{EiFe}$^-$\cite{Mann}, which provide the modern
theoretical framework for the description of the properties and
decays of hadrons containing a heavy quark. For a more detailed
description of this subject, the reader is referred to the review
articles in Refs.~24--30.

\subsection{The Physical Picture}

There are several reasons why the strong interactions of hadrons
containing heavy quarks are easier to understand than those of
hadrons containing only light quarks. The first is asymptotic
freedom, the fact that the effective coupling constant of QCD becomes
weak in processes with a large momentum transfer, corresponding to
interactions at short distance scales~\cite{Gros,Poli}. At large
distances, on the other hand, the coupling becomes strong, leading to
non-perturbative phenomena such as the confinement of quarks and
gluons on a length scale $R_{\rm had}\sim 1/\Lambda_{\rm QCD}\sim
1$~fm, which determines the size of hadrons. Roughly speaking,
$\Lambda_{\rm QCD}\sim 0.2$ GeV is the energy scale that separates
the regions of large and small coupling constant. When the mass of a
quark $Q$ is much larger than this scale, $m_Q\gg\Lambda_{\rm QCD}$,
it is called a heavy quark. The quarks of the Standard Model fall
naturally into two classes: up, down and strange are light quarks,
whereas charm, bottom and top are heavy quarks.\footnote{Ironically,
the top quark is of no relevance to our discussion here, since it is
too heavy to form hadronic bound states before it decays.}
For heavy quarks, the effective coupling constant $\alpha_s(m_Q)$ is
small, implying that on length scales comparable to the Compton
wavelength $\lambda_Q\sim 1/m_Q$ the strong interactions are
perturbative and much like the electromagnetic interactions. In fact,
the quarkonium systems $(\bar QQ)$, whose size is of order
$\lambda_Q/\alpha_s(m_Q)\ll R_{\rm had}$, are very much
hydrogen-like.

Systems composed of a heavy quark and other light constituents are
more complicated. The size of such systems is determined by $R_{\rm
had}$, and the typical momenta exchanged between the heavy and light
constituents are of order $\Lambda_{\rm QCD}$. The heavy quark is
surrounded by a most complicated, strongly interacting cloud of light
quarks, antiquarks and gluons. In this case it is the fact that
$\lambda_Q\ll R_{\rm had}$, i.e.\ that the Compton wavelength of the
heavy quark is much smaller than the size of the hadron, which leads
to simplifications. To resolve the quantum numbers of the heavy quark
would require a hard probe; the soft gluons exchanged between the
heavy quark and the light constituents can only resolve distances
much larger than $\lambda_Q$. Therefore, the light degrees of freedom
are blind to the flavour (mass) and spin orientation of the heavy
quark. They experience only its colour field, which extends over
large distances because of confinement. In the rest frame of the
heavy quark, it is in fact only the electric colour field that is
important; relativistic effects such as colour magnetism vanish as
$m_Q\to\infty$. Since the heavy-quark spin participates in
interactions only through such relativistic effects, it decouples.
That the heavy-quark mass becomes irrelevant can be seen as follows:
As $m_Q\to\infty$, the heavy quark and the hadron that contains it
have the same velocity. In the rest frame of the hadron, the heavy
quark is at rest, too. The wave function of the light constituents
follows from a solution of the field equations of QCD subject to the
boundary condition of a static triplet source of colour at the
location of the heavy quark. This boundary condition is independent
of $m_Q$, and so is the solution for the configuration of the light
constituents.

It follows that, in the limit $m_Q\to\infty$, hadronic systems which
differ only in the flavour or spin quantum numbers of the heavy quark
have the same configuration of their light degrees of
freedom~\cite{Shu1}$^-$\cite{Isgu}. Although this observation still
does not allow us to calculate what this configuration is, it
provides relations between the properties of such particles as the
heavy mesons $B$, $D$, $B^*$ and $D^*$, or the heavy baryons
$\Lambda_b$ and $\Lambda_c$ (to the extent that corrections to the
infinite quark-mass limit are small in these systems). These
relations result from some approximate symmetries of the effective
strong interactions of heavy quarks at low energies. The
configuration of light degrees of freedom in a hadron containing a
single heavy quark with velocity $v$ does not change if this quark is
replaced by another heavy quark with different flavour or spin, but
with the same velocity. Both heavy quarks lead to the same static
colour field. For $N_h$ heavy-quark flavours, there is thus an SU$(2
N_h)$ spin-flavour symmetry group, under which the effective strong
interactions are invariant. These symmetries are in close
correspondence to familiar properties of atoms. The flavour symmetry
is analogous to the fact that different isotopes have the same
chemistry, since to good approximation the wave function of the
electrons is independent of the mass of the nucleus. The electrons
only see the total nuclear charge. The spin symmetry is analogous to
the fact that the hyperfine levels in atoms are nearly degenerate.
The nuclear spin decouples in the limit $m_e/m_N\to 0$.

Heavy-quark symmetry is an approximate symmetry, and corrections
arise since the quark masses are not infinite. In many respects, it
is complementary to chiral symmetry, which arises in the opposite
limit of small quark masses. There is an important distinction,
however. Whereas chiral symmetry is a symmetry of the QCD Lagrangian
in the limit of vanishing quark masses, heavy-quark symmetry is not a
symmetry of the Lagrangian (not even an approximate one), but rather
a symmetry of an effective theory, which is a good approximation of
QCD in a certain kinematic region. It is realized only in systems in
which a heavy quark interacts predominantly by the exchange of soft
gluons. In such systems the heavy quark is almost on-shell; its
momentum fluctuates around the mass shell by an amount of order
$\Lambda_{\rm QCD}$. The corresponding fluctuations in the velocity
of the heavy quark vanish as $\Lambda_{\rm QCD}/m_Q\to 0$. The
velocity becomes a conserved quantity and is no longer a dynamical
degree of freedom~\cite{Geor}. Nevertheless, results derived on the
basis of heavy-quark symmetry are model-independent consequences of
QCD in a well-defined limit. The symmetry-breaking corrections can be
studied in a systematic way. To this end, it is however necessary to
cast the QCD Lagrangian for a heavy quark,
\begin{equation}\label{QCDLag}
   {\cal L}_Q = \bar Q\,(i\rlap{\,/}D - m_Q)\,Q \,,
\end{equation}
into a form suitable for taking the limit $m_Q\to\infty$.

\subsection{Heavy-Quark Effective Theory}

The effects of a very heavy particle often become irrelevant at low
energies. It is then useful to construct a low-energy effective
theory, in which this heavy particle no longer appears. Eventually,
this effective theory will be easier to deal with than the full
theory. A familiar example is Fermi's theory of the weak
interactions. For the description of the weak decays of hadrons, the
weak interactions can be approximated by point-like four-fermion
couplings, governed by a dimensionful coupling constant $G_F$
[cf.~(\ref{LFermi})]. The effects of the intermediate vector bosons,
$W$ and $Z$, can only be resolved at energies much larger than the
hadron masses.

The process of removing the degrees of freedom of a heavy particle
involves the following steps~\cite{SVZ1}$^-$\cite{Polc}: one first
identifies the heavy-particle fields and ``integrates them out'' in
the generating functional of the Green functions of the theory. This
is possible since at low energies the heavy particle does not appear
as an external state. However, although the action of the full theory
is usually a local one, what results after this first step is a
non-local effective action. The non-locality is related to the fact
that in the full theory the heavy particle with mass $M$ can appear
in virtual processes and propagate over a short but finite distance
$\Delta x\sim 1/M$. Thus, a second step is required to obtain a local
effective Lagrangian: the non-local effective action is rewritten as
an infinite series of local terms in an Operator Product Expansion
(OPE)~\cite{Wils,Zimm}. Roughly speaking, this corresponds to an
expansion in powers of $1/M$. It is in this step that the short- and
long-distance physics is disentangled. The long-distance physics
corresponds to interactions at low energies and is the same in the
full and the effective theory. But short-distance effects arising
from quantum corrections involving large virtual momenta (of order
$M$) are not reproduced in the effective theory, once the heavy
particle has been integrated out. In a third step, they have to be
added in a perturbative way using renormalization-group techniques.
These short-distance effects lead to a renormalization of the
coefficients of the local operators in the effective Lagrangian. An
example is the effective Lagrangian for non-leptonic weak decays, in
which radiative corrections from hard gluons with virtual momenta in
the range between $m_W$ and some renormalization scale $\mu\sim
1$~GeV give rise to Wilson coefficients, which renormalize the local
four-fermion interactions~\cite{AltM}$^-$\cite{Gilm}.

\begin{figure}[htb]
   \epsfysize=7cm
   \centerline{\epsffile{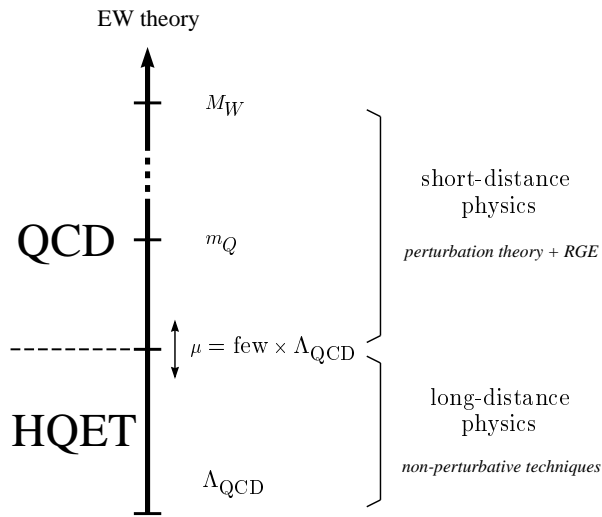}}
\caption{\label{fig:magic}
Philosophy of the heavy-quark effective theory.}
\end{figure}

The heavy-quark effective theory (HQET) is constructed to provide a
simplified description of processes where a heavy quark interacts
with light degrees of freedom predominantly by the exchange of soft
gluons~\cite{EiFe}$^-$\cite{Mann}. Clearly, $m_Q$ is the high-energy
scale in this case, and $\Lambda_{\rm QCD}$ is the scale of the
hadronic physics we are interested in. The situation is illustrated
in Fig.~\ref{fig:magic}. At short distances, i.e.\ for energy scales
larger than the heavy-quark mass, the physics is perturbative and
described by conventional QCD. For mass scales much below the
heavy-quark mass, the physics is complicated and non-perturbative
because of confinement. Our goal is to obtain a simplified
description in this region using an effective field theory. To
separate short- and long-distance effects, we introduce a separation
scale $\mu$ such that $\Lambda_{\rm QCD}\ll\mu\ll m_Q$. The HQET will
be constructed in such a way that it is identical to QCD in the
long-distance region, i.e.\ for scales below $\mu$. In the
short-distance region, the effective theory is incomplete, since some
high-momentum modes have been integrated out from the full theory.
The fact that the physics must be independent of the arbitrary scale
$\mu$ allows us to derive renormalization-group equations, which can
be employed to deal with the short-distance effects in an efficient
way.

Compared with most effective theories, in which the degrees of
freedom of a heavy particle are removed completely from the
low-energy theory, the HQET is special in that its purpose is to
describe the properties and decays of hadrons which do contain a
heavy quark. Hence, it is not possible to remove the heavy quark
completely from the effective theory. What is possible is to
integrate out the ``small components'' in the full heavy-quark
spinor, which describe the fluctuations around the mass shell.

The starting point in the construction of the HQET is the observation
that a heavy quark bound inside a hadron moves more or less with the
hadron's velocity $v$, and is almost on-shell. Its momentum can be
written as
\begin{equation}\label{kresdef}
   p_Q^\mu = m_Q v^\mu + k^\mu \,,
\end{equation}
where the components of the so-called residual momentum $k$ are much
smaller than $m_Q$. Note that $v$ is a four-velocity, so that
$v^2=1$. Interactions of the heavy quark with light degrees of
freedom change the residual momentum by an amount of order $\Delta
k\sim\Lambda_{\rm QCD}$, but the corresponding changes in the
heavy-quark velocity vanish as $\Lambda_{\rm QCD}/m_Q\to 0$. In this
situation, it is appropriate to introduce large- and small-component
fields, $h_v$ and $H_v$, by
\begin{equation}\label{hvHvdef}
   h_v(x) = e^{i m_Q v\cdot x}\,P_+\,Q(x) \,, \qquad
   H_v(x) = e^{i m_Q v\cdot x}\,P_-\,Q(x) \,,
\end{equation}
where $P_+$ and $P_-$ are projection operators defined as
\begin{equation}
   P_\pm = {1\pm\rlap/v\over 2} \,.
\end{equation}
It follows that
\begin{equation}\label{redef}
   Q(x) = e^{-i m_Q v\cdot x}\,[ h_v(x) + H_v(x) ] \,.
\end{equation}
Because of the projection operators, the new fields satisfy
$\rlap/v\,h_v=h_v$ and $\rlap/v\,H_v=-H_v$. In the rest frame, i.e.\
for $v^\mu=(1,0,0,0)$, $h_v$ corresponds to the upper two components
of $Q$, while $H_v$ corresponds to the lower ones. Whereas $h_v$
annihilates a heavy quark with velocity $v$, $H_v$ creates a heavy
antiquark with velocity $v$.

\begin{figure}[htb]
   \epsfxsize=7cm
   \centerline{\epsffile{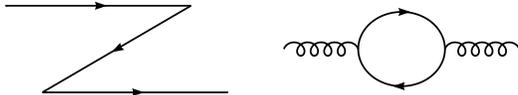}}
\caption{\label{fig:3.1}
Virtual fluctuations involving pair creation of heavy quarks. Time
flows to the right.}
\end{figure}

In terms of the new fields, the QCD Lagrangian (\ref{QCDLag}) for a
heavy quark takes the form
\begin{equation}\label{Lhchi}
   {\cal L}_Q = \bar h_v\,i v\!\cdot\!D\,h_v
   - \bar H_v\,(i v\!\cdot\!D + 2 m_Q)\,H_v
   + \bar h_v\,i\rlap{\,/}D_\perp H_v
   + \bar H_v\,i\rlap{\,/}D_\perp h_v \,,
\end{equation}
where $D_\perp^\mu = D^\mu - v^\mu\,v\cdot D$ is orthogonal to the
heavy-quark velocity: $v\cdot D_\perp=0$. In the rest frame,
$D_\perp^\mu=(0,\vec D\,)$ contains the spatial components of the
covariant derivative. From (\ref{Lhchi}), it is apparent that $h_v$
describes massless degrees of freedom, whereas $H_v$ corresponds to
fluctuations with twice the heavy-quark mass. These are the heavy
degrees of freedom that will be eliminated in the construction of the
effective theory. The fields are mixed by the presence of the third
and fourth terms, which describe pair creation or annihilation of
heavy quarks and antiquarks. As shown in the first diagram in
Fig.~\ref{fig:3.1}, in a virtual process, a heavy quark propagating
forward in time can turn into an antiquark propagating backward in
time, and then turn back into a quark. The energy of the intermediate
quantum state $h h\bar H$ is larger than the energy of the incoming
heavy quark by at least $2 m_Q$. Because of this large energy gap,
the virtual quantum fluctuation can only propagate over a short
distance $\Delta x\sim 1/m_Q$. On hadronic scales set by $R_{\rm
had}=1/\Lambda_{\rm QCD}$, the process essentially looks like a local
interaction of the form
\begin{equation}
   \bar h_v\,i\rlap{\,/}D_\perp\,{1\over 2 m_Q}\,
   i\rlap{\,/}D_\perp h_v \,,
\end{equation}
where we have simply replaced the propagator for $H_v$ by $1/2 m_Q$.
A more correct treatment is to integrate out the small-component
field $H_v$, thereby deriving a non-local effective action for the
large-component field $h_v$, which can then be expanded in terms of
local operators. Before doing this, let us mention a second type of
virtual corrections involving pair creation, namely heavy-quark
loops. An example is shown in the second diagram in
Fig.~\ref{fig:3.1}. Heavy-quark loops cannot be described in terms of
the effective fields $h_v$ and $H_v$, since the quark velocities
inside a loop are not conserved and are in no way related to hadron
velocities. However, such short-distance processes are proportional
to the small coupling constant $\alpha_s(m_Q)$ and can be calculated
in perturbation theory. They lead to corrections that are added onto
the low-energy effective theory in the renormalization procedure.

On a classical level, the heavy degrees of freedom represented by
$H_v$ can be eliminated using the equation of motion. Taking the
variation of the Lagrangian with respect to the field $\bar H_v$, we
obtain
\begin{equation}
   (i v\!\cdot\!D + 2 m_Q)\,H_v = i\rlap{\,/}D_\perp h_v \,.
\end{equation}
This equation can formally be solved to give
\begin{equation}\label{Hfield}
   H_v = {1\over 2 m_Q + i v\!\cdot\!D}\,
   i\rlap{\,/}D_\perp h_v \,,
\end{equation}
showing that the small-component field $H_v$ is indeed of order
$1/m_Q$. We can now insert this solution into (\ref{Lhchi}) to obtain
the ``non-local effective Lagrangian''
\begin{equation}\label{Lnonloc}
   {\cal L}_{\rm eff} = \bar h_v\,i v\!\cdot\!D\,h_v
   + \bar h_v\,i\rlap{\,/}D_\perp\,{1\over 2 m_Q+i v\!\cdot\!D}\,
   i\rlap{\,/}D_\perp h_v \,.
\end{equation}
Clearly, the second term corresponds to the first class of virtual
processes shown in Fig.~\ref{fig:3.1}.

It is possible to derive this Lagrangian in a more elegant way by
manipulating the generating functional for QCD Green functions
containing heavy-quark fields~\cite{Mann}. To this end, one starts
from the field redefinition (\ref{redef}) and couples the
large-component fields $h_v$ to external sources $\rho_v$. Green
functions with an arbitrary number of $h_v$ fields can be constructed
by taking derivatives with respect to $\rho_v$. No sources are needed
for the heavy degrees of freedom represented by $H_v$. The functional
integral over these fields is Gaussian and can be performed
explicitly, leading to the effective action
\begin{equation}\label{SeffMRR}
   S_{\rm eff} = \int\!{\rm d}^4 x\,{\cal L}_{\rm eff}
   - i \ln\Delta \,,
\end{equation}
with ${\cal L}_{\rm eff}$ as given in (\ref{Lnonloc}). The
appearance
of the logarithm of the determinant
\begin{equation}
   \Delta = \exp\bigg( {1\over 2}\,{\rm Tr}\,
   \ln\big[ 2 m_Q + i v\!\cdot\!D - i\eta \big] \bigg)
\end{equation}
is a quantum effect not present in the classical derivation presented
above. However, in this case the determinant can be regulated in a
gauge-invariant way, and by choosing the gauge $v\cdot A=0$ one can
show that $\ln\Delta$ is just an irrelevant
constant~\cite{Mann,Soto}.

\begin{figure}[htb]
   \epsfysize=3.5cm
   \centerline{\epsffile{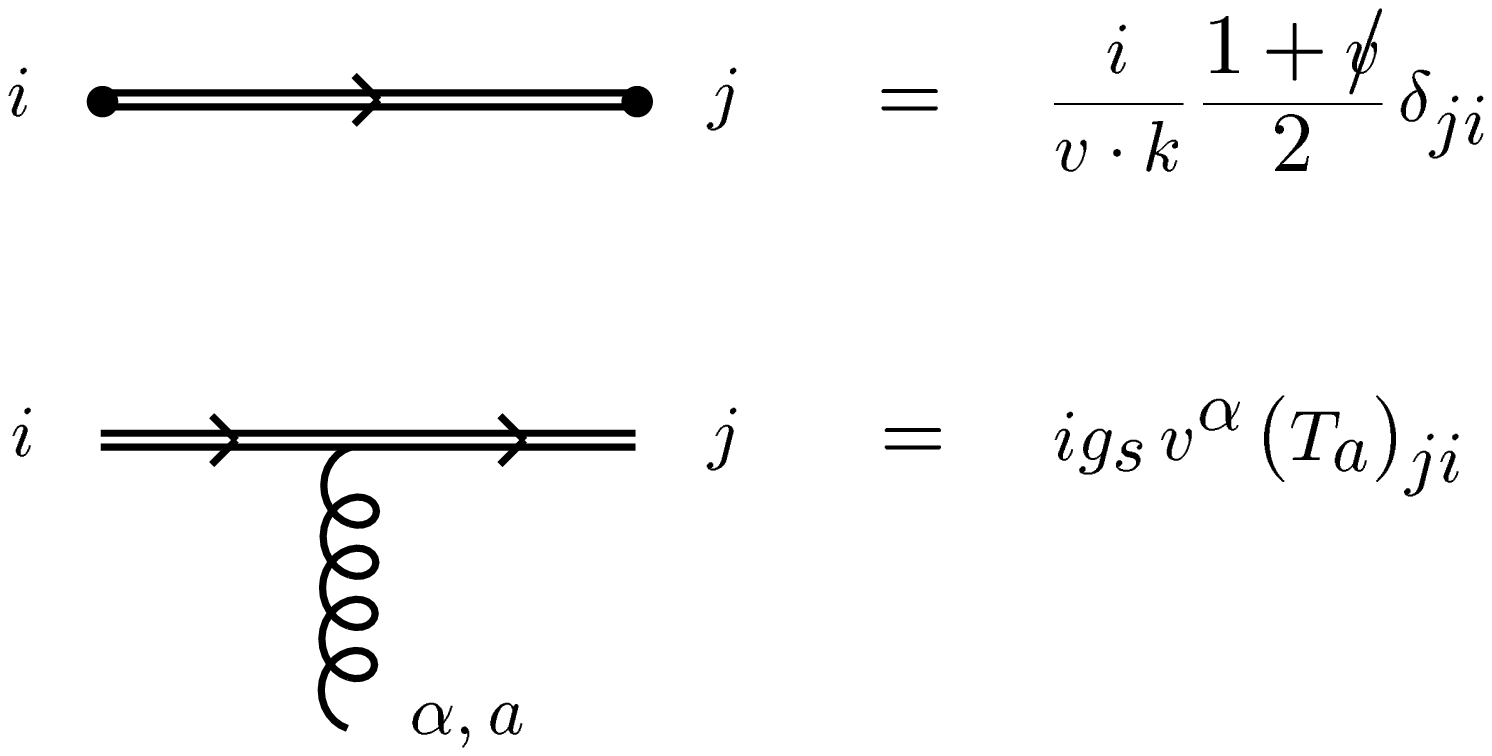}}
\caption{\label{fig:3.2}
Feynman rules of the HQET ($i,j$ and $a$ are colour indices). A heavy
quark with velocity $v$ is represented by a double line. The residual
momentum $k$ is defined in (\protect\ref{kresdef}).}
\end{figure}

Because of the phase factor in (\ref{redef}), the $x$ dependence of
the effective heavy-quark field $h_v$ is weak. In momentum space,
derivatives acting on $h_v$ produce powers of the residual momentum
$k$, which is much smaller than $m_Q$. Hence, the non-local effective
Lagrangian (\ref{Lnonloc}) allows for a derivative expansion:
\begin{equation}
   {\cal L}_{\rm eff} = \bar h_v\,i v\!\cdot\!D\,h_v
   + {1\over 2 m_Q}\,\sum_{n=0}^\infty\,
   \bar h_v\,i\rlap{\,/}D_\perp\,\bigg( -{i v\cdot D\over 2 m_Q}
   \bigg)^n\,i\rlap{\,/}D_\perp h_v \,.
\end{equation}
Taking into account that $h_v$ contains a $P_+$ projection operator,
and using the identity
\begin{equation}\label{pplusid}
   P_+\,i\rlap{\,/}D_\perp\,i\rlap{\,/}D_\perp P_+
   = P_+\,\bigg[ (i D_\perp)^2 + {g_s\over 2}\,
   \sigma_{\mu\nu }\,G^{\mu\nu } \bigg]\,P_+ \,,
\end{equation}
where $i[D^\mu,D^\nu]=g_s\,G^{\mu\nu}$ is the gluon field-strength
tensor, one finds that~\cite{EiH1,FGL}
\begin{equation}\label{Lsubl}
   {\cal L}_{\rm eff} = \bar h_v\,i v\!\cdot\!D\,h_v
   + {1\over 2 m_Q}\,\bar h_v\,(i D_\perp)^2\,h_v
   + {g_s\over 4 m_Q}\,\bar h_v\,\sigma_{\mu\nu}\,
   G^{\mu\nu}\,h_v + O(1/m_Q^2) \,.
\end{equation}
In the limit $m_Q\to\infty$, only the first terms remains:
\begin{equation}\label{Leff}
   {\cal L}_\infty = \bar h_v\,i v\!\cdot\!D\,h_v \,.
\end{equation}
This is the effective Lagrangian of the HQET. It gives rise to the
Feynman rules shown in Fig.~\ref{fig:3.2}.

Let us take a moment to study the symmetries of this
Lagrangian~\cite{Geor}. Since there appear no Dirac matrices,
interactions of the heavy quark with gluons leave its spin unchanged.
Associated with this is an SU(2) symmetry group, under which
${\cal L}_\infty$ is invariant. The action of this symmetry on the
heavy-quark fields becomes most transparent in the rest frame, where
the generators $S^i$ of SU(2) can be chosen as
\begin{equation}\label{Si}
   S^i = {1\over 2} \left( \begin{array}{cc}
                           \sigma^i ~&~ 0 \\
                           0 ~&~ \sigma^i \end{array} \right) \,;
   \qquad [S^i,S^j] = i \epsilon^{ijk} S^k \,.
\end{equation}
Here $\sigma^i$ are the Pauli matrices. An infinitesimal SU(2)
transformation $h_v\to (1 + i\vec\epsilon \cdot\vec S\,)\,h_v$ leaves
the Lagrangian invariant:
\begin{equation}\label{SU2tr}
   \delta{\cal L}_\infty = \bar h_v\,
   [i v\!\cdot\! D,i \vec\epsilon\cdot\vec S\,]\,h_v = 0 \,.
\end{equation}
Another symmetry of the HQET arises since the mass of the heavy quark
does not appear in the effective Lagrangian. For $N_h$ heavy quarks
moving at the same velocity, eq.~(\ref{Leff}) can be extended by
writing
\begin{equation}\label{Leff2}
   {\cal L}_\infty
   = \sum_{i=1}^{N_h} \bar h_v^i\,i v\!\cdot\! D\,h_v^i \,.
\end{equation}
This is invariant under rotations in flavour space. When combined
with the spin symmetry, the symmetry group is promoted to SU$(2
N_h)$. This is the heavy-quark spin-flavour
symmetry~\cite{Isgu,Geor}.
Its physical content is that, in the limit $m_Q\to\infty$, the strong
interactions of a heavy quark become independent of its mass and
spin.

Consider now the operators appearing at order $1/m_Q$ in the
effective Lagrangian (\ref{Lsubl}). They are easiest to identify in
the rest frame. The first operator,
\begin{equation}\label{Okin}
   {\cal O}_{\rm kin} = {1\over 2 m_Q}\,\bar h_v\,(i D_\perp)^2\,
   h_v \to - {1\over 2 m_Q}\,\bar h_v\,(i \vec D\,)^2\,h_v \,,
\end{equation}
is the gauge-covariant extension of the kinetic energy arising from
the residual motion of the heavy quark. The second operator is the
non-Abelian analogue of the Pauli interaction, which describes the
colour-magnetic coupling of the heavy-quark spin to the gluon field:
\begin{equation}\label{Omag}
   {\cal O}_{\rm mag} = {g_s\over 4 m_Q}\,\bar h_v\,
   \sigma_{\mu\nu}\,G^{\mu\nu}\,h_v \to
   - {g_s\over m_Q}\,\bar h_v\,\vec S\!\cdot\!\vec B_c\,h_v \,.
\end{equation}
Here $\vec S$ is the spin operator defined in (\ref{Si}), and $B_c^i
= -\frac{1}{2}\epsilon^{ijk} G^{jk}$ are the components of the
colour-magnetic field. The chromo-magnetic interaction is a
relativistic effect, which scales like $1/m_Q$. This is the origin of
the heavy-quark spin symmetry.

\subsection{The Residual Mass Term and the Definition of the
Heavy-Quark Mass}

The choice of the expansion parameter in the HQET, i.e.\ the
definition of the heavy-quark mass $m_Q$, deserves some comments. In
the derivation presented earlier in this section, we chose $m_Q$ to
be the ``mass in the Lagrangian'', and using this parameter in the
phase redefinition in (\ref{redef}) we obtained the effective
Lagrangian (\ref{Leff}), in which the heavy-quark mass no longer
appears. However, this treatment has its subtleties. The symmetries
of the HQET allow a ``residual mass'' $\delta m$ for the heavy quark,
provided that $\delta m$ is of order $\Lambda_{\rm QCD}$ and is the
same for all heavy-quark flavours. Even if we arrange that such a
mass term is not present at the tree level, it will in general be
induced by quantum corrections. (This is unavoidable if the theory is
regulated with a dimensionful cutoff.) Therefore, instead of
(\ref{Leff}) we should write the effective Lagrangian in the more
general form~\cite{FNL}
\begin{equation}
   {\cal L}_\infty = \bar h_v\,iv\!\cdot\!D\,h_v
   - \delta m\,\bar h_v h_v \,.
\end{equation}
If we redefine the expansion parameter according to $m_Q\to
m_Q+\Delta m$, the residual mass changes in the opposite way: $\delta
m\to\delta m-\Delta m$. This implies that there is a unique choice of
the expansion parameter $m_Q$ such that $\delta m=0$. Requiring
$\delta m=0$, as it is usually done implicitly in the HQET, defines a
heavy-quark mass, which in perturbation theory coincides with the
pole mass~\cite{Tarr}. This, in turn, defines for each heavy hadron
$H_Q$ a parameter $\bar\Lambda$ (sometimes called the ``binding
energy'') through
\begin{equation}
   \bar\Lambda = (m_{H_Q} - m_Q)\Big|_{m_Q\to\infty} \,.
\label{Lbdef}
\end{equation}
If one prefers to work with another choice of the expansion
parameter, the values of non-perturbative parameters such as
$\bar\Lambda$ change, but at the same time one has to include the
residual mass term in the HQET Lagrangian. It can be shown that the
various parameters depending on the definition of $m_Q$ enter the
predictions for physical quantities in such a way that the results
are independent of the particular choice adopted~\cite{FNL}.

There is one more subtlety hidden in the above discussion. The
quantities $m_Q$, $\bar\Lambda$ and $\delta m$ are non-perturbative
parameters of the HQET, which have a similar status as the vacuum
condensates in QCD phenomenology~\cite{SVZ}. These parameters cannot
be defined unambiguously in perturbation theory. The reason lies in
the divergent behaviour of perturbative expansions in large orders,
which is associated with the existence of singularities along the
real axis in the Borel plane, the so-called
renormalons~\cite{tHof}$^-$\cite{Muel}. For instance, the
perturbation series which relates the pole mass $m_Q$ of a heavy
quark to its bare mass,
\begin{equation}
   m_Q = m_Q^{\rm bare}\,\Big\{ 1 + c_1\,\alpha_s(m_Q)
   + c_2\,\alpha_s^2(m_Q) + \dots + c_n\,\alpha_s^n(m_Q)
   + \dots \Big\} \,,
\end{equation}
contains numerical coefficients $c_n$ that grow as $n!$ for large
$n$, rendering the series divergent and not Borel
summable~\cite{BBren,Bigiren}. The best one can achieve is to
truncate the perturbation series at the minimal term, but this leads
to an unavoidable arbitrariness of order $\Delta m_Q\sim\Lambda_{\rm
QCD}$ (the size of the minimal term) in the value of the pole mass.
This observation, which at first sight seems a serious problem for
QCD phenomenology, should not come as a surprise. We know that
because of confinement quarks do not appear as physical states in
nature. Hence, there is no unique way to define their on-shell
properties such as a pole mass. In view of this, it is actually
remarkable that QCD perturbation theory ``knows'' about its
incompleteness and indicates, through the appearance of renormalon
singularities, the presence of non-perturbative effects. One must
first specify a scheme how to truncate the QCD perturbation series
before non-perturbative statements such as $\delta m=0$ become
meaningful, and hence before non-perturbative parameters such as
$m_Q$ and $\bar\Lambda$ become well-defined quantities. The actual
values of these parameters will depend on this scheme.

We stress that the ``renormalon ambiguities'' are not a conceptual
problem for the heavy-quark expansion. In fact, it can be shown quite
generally that these ambiguities cancel in all predictions for
physical observables~\cite{Chris}$^-$\cite{LMS}. The way the
cancellations occur is intricate, however. The generic structure of
the heavy-quark expansion for an observable is of the form:
\begin{equation}
   \mbox{Observable} \sim C[\alpha_s(m_Q)]\,\bigg( 1
   + {\Lambda\over m_Q} + \dots \bigg) \,,
\end{equation}
where $C[\alpha_s(m_Q)]$ represents a perturbative coefficient
function, and $\Lambda$ is a dimensionful non-perturbative parameter.
The truncation of the perturbation series defining the coefficient
function leads to an arbitrariness of order $\Lambda_{\rm QCD}/m_Q$,
which cancels against a corresponding arbitrariness of order
$\Lambda_{\rm QCD}$ in the definition of the non-perturbative
parameter $\Lambda$.

The renormalon problem poses itself when one imagines to apply
perturbation theory in very high orders. In practise, the
perturbative coefficients are known to finite order in $\alpha_s$ (at
best to two-loop accuracy), and to be consistent one should use them
in connection with the pole mass (and $\bar\Lambda$ etc.) defined to
the same order.

\subsection{Spectroscopic Implications}

The spin-flavour symmetry leads to many interesting relations between
the properties of hadrons containing a heavy quark. The most direct
consequences concern the spectroscopy of such states~\cite{IsWi}. In
the limit $m_Q\to\infty$, the spin of the heavy quark and the total
angular momentum $j$ of the light degrees of freedom are separately
conserved by the strong interactions. Because of heavy-quark
symmetry, the dynamics is independent of the spin and mass of the
heavy quark. Hadronic states can thus be classified by the quantum
numbers (flavour, spin, parity, etc.) of their light degrees of
freedom~\cite{AFal}. The spin symmetry predicts that, for fixed
$j\neq
0$, there is a doublet of degenerate states with total spin
$J=j\pm\frac{1}{2}$. The flavour symmetry relates the properties of
states with different heavy-quark flavour.

In general, the mass of a hadron $H_Q$ containing a heavy quark $Q$
obeys an expansion of the form
\begin{equation}\label{massexp}
   m_{H_Q} = m_Q + \bar\Lambda + {\Delta m^2\over 2 m_Q}
   + O(1/m_Q^2) \,.
\end{equation}
The parameter $\bar\Lambda$ represents contributions arising from
terms in the Lagrangian that are independent of the heavy-quark
mass~\cite{FNL}, whereas the quantity $\Delta m^2$ originates from
the terms of order $1/m_Q$ in the effective Lagrangian of the HQET.
For the ground-state pseudoscalar and vector mesons, one can
parame\-trize the contributions from the kinetic energy and the
chromo-magnetic interaction in terms of two quantities $\lambda_1$
and $\lambda_2$, in such a way that~\cite{FaNe}
\begin{equation}\label{FNrela}
   \Delta m^2 = -\lambda_1 + 2 \Big[ J(J+1) - \textstyle{3\over 2}
   \Big]\,\lambda_2 \,.
\end{equation}
The hadronic parameters $\bar\Lambda$, $\lambda_1$ and $\lambda_2$
are independent of $m_Q$. They characterize the properties of the
light constituents.

Consider, as a first example, the SU(3) mass splittings for heavy
mesons. The heavy-quark expansion predicts that
\begin{eqnarray}
   m_{B_S} - m_{B_d} &=& \bar\Lambda_s - \bar\Lambda_d
    + O(1/m_b) \,, \nonumber\\
   m_{D_S} - m_{D_d} &=& \bar\Lambda_s - \bar\Lambda_d
    + O(1/m_c) \,,
\end{eqnarray}
where we have indicated that the value of the parameter $\bar\Lambda$
depends on the flavour of the light quark. Thus, to the extent that
the charm and bottom quarks can both be considered sufficiently
heavy, the mass splittings should be similar in the two systems. This
prediction is confirmed experimentally, since~\cite{Joe}
\begin{eqnarray}
   m_{B_S} - m_{B_d} &=& (90\pm 3)~\mbox{MeV} \,, \nonumber\\
   m_{D_S} - m_{D_d} &=& (99\pm 1)~\mbox{MeV} \,.
\end{eqnarray}

As a second example, consider the spin splittings between the
ground-state pseudoscalar ($J=0$) and vector ($J=1$) mesons, which
are the members of the spin-doublet with $j=\frac{1}{2}$. From
(\ref{massexp}) and (\ref{FNrela}), it follows that
\begin{eqnarray}
   m_{B^*}^2 - m_B^2 &=& 4\lambda_2 + O(1/m_b) \,, \nonumber\\
   m_{D^*}^2 - m_D^2 &=& 4\lambda_2 + O(1/m_c) \,.
\end{eqnarray}
The data are compatible with this:
\begin{eqnarray}\label{VPexp}
   m_{B^*}^2 - m_B^2 &\approx& 0.49~{\rm GeV}^2 \,, \nonumber\\
   m_{D^*}^2 - m_D^2 &\approx& 0.55~{\rm GeV}^2 \,.
\end{eqnarray}
Assuming that the $B$ system is close to the heavy-quark limit, we
obtain the value
\begin{equation}
   \lambda_2\approx 0.12~\mbox{GeV}^2
\label{lam2val}
\end{equation}
for one of the hadronic parameters in (\ref{FNrela}). This quantity
plays an important role in the phenomenology of inclusive decays of
heavy hadrons.

A third example is provided by the mass splittings between the
ground-state mesons and baryons containing a heavy quark. The HQET
predicts that
\begin{eqnarray}\label{barmes}
   m_{\Lambda_b} - m_B &=& \bar\Lambda_{\rm baryon}
    - \bar\Lambda_{\rm meson} + O(1/m_b) \,, \nonumber\\
   m_{\Lambda_c} - m_D &=& \bar\Lambda_{\rm baryon}
    - \bar\Lambda_{\rm meson} + O(1/m_c) \,.
\end{eqnarray}
This is again consistent with the experimental results
\begin{eqnarray}
   m_{\Lambda_b} - m_B &=& (346\pm 6)~\mbox{MeV} \,, \nonumber\\
   m_{\Lambda_c} - m_D &=& (416\pm 1)~\mbox{MeV} \,,
\end{eqnarray}
although in this case the data indicate sizeable symmetry-breaking
corrections. For the mass of the $\Lambda_b$ baryon, we have used the
value
\begin{equation}\label{Lbmass}
   m_{\Lambda_b} = (5625\pm 6)~\mbox{MeV} \,,
\end{equation}
which is obtained by averaging the result~\cite{Joe} $m_{\Lambda_b}=
(5639\pm 15)$~MeV with the value $m_{\Lambda_b}=(5623\pm 5\pm 4)$~MeV
reported by the CDF Collaboration~\cite{CDFmass}. The dominant
correction to the relations (\ref{barmes}) comes from the
contribution of the chromo-magnetic interaction to the masses of the
heavy mesons,\footnote{Because of the spin symmetry, there is no such
contribution to the masses of the $\Lambda_Q$ baryons.} which adds a
term $3\lambda_2/2 m_Q$ on the right-hand side. Including this term,
we obtain the refined prediction that the two quantities
\begin{eqnarray}
   m_{\Lambda_b} - m_B - {3\lambda_2\over 2 m_B}
   &=& (312\pm 6)~\mbox{MeV} \,, \nonumber\\
   m_{\Lambda_c} - m_D - {3\lambda_2\over 2 m_D}
   &=& (320\pm 1)~\mbox{MeV}
\end{eqnarray}
should be close to each other. This is clearly satisfied by the data.

The mass formula (\ref{massexp}) can also be used to derive
information on the heavy-quark masses from the observed hadron
masses. Introducing the ``spin-averaged'' meson masses
$\overline{m}_B=\frac{1}{4}\,(m_B+3 m_{B^*})\approx 5.31$~GeV and
$\overline{m}_D=\frac{1}{4}\,(m_D+3 m_{D^*})\approx 1.97$~GeV, we
find that
\begin{equation}\label{mbmc}
   m_b-m_c = (\overline{m}_B-\overline{m}_D)\,\bigg\{
   1 - {\lambda_1\over 2\overline{m}_B\overline{m}_D}
   + O(1/m_Q^3) \bigg\} \,.
\end{equation}
Using theoretical estimates for the parameter $\lambda_1$, which lie
in the range~\cite{ElSh}$^-$\cite{Fazi}
\begin{equation}\label{lam1}
   \lambda_1 = -(0.3\pm 0.2)~\mbox{GeV}^2 \,,
\end{equation}
this relation leads to
\begin{equation}\label{mbmcval}
   m_b - m_c = (3.39\pm 0.03\pm 0.03)~\mbox{GeV} \,,
\label{mQdif}
\end{equation}
where the first error reflects the uncertainty in the value of
$\lambda_1$, and the second one takes into account unknown
higher-order corrections. The fact that the difference $(m_b-m_c)$ is
determined rather precisely becomes important in the analysis of
inclusive decays of heavy hadrons.

\section{Exclusive Semileptonic Decays}
\label{sec:3}

\begin{figure}[htb]
   \epsfxsize=6.5cm
   \centerline{\epsffile{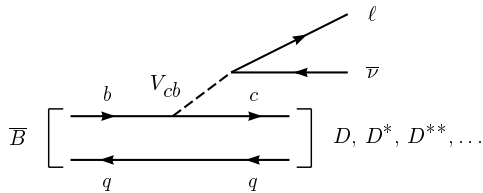}}
\caption{\label{fig:1}
Semileptonic decays of $B$ mesons.}
\end{figure}

Semileptonic decays of $B$ mesons have received a lot of attention in
recent years. The decay channel $\bar B\to D^*\ell\,\bar\nu$ has the
largest branching fraction of all $B$-meson decay modes. From a
theoretical point of view, semileptonic decays are simple enough to
allow for a reliable, quantitative description. The analysis of these
decays provides much information about the strong forces that bind
the quarks and gluons into hadrons. Schematically, a semileptonic
decay process is shown in Fig.~\ref{fig:1}. The strength of the $b\to
c$ transition vertex is governed by the element $V_{cb}$ of the CKM
matrix. The parameters of this matrix are fundamental parameters of
the Standard Model. A primary goal of the study of semileptonic
decays of $B$ mesons is to extract with high precision the values of
$|V_{cb}|$ and $|V_{ub}|$. We will now discuss the theoretical basis
of such analyses.

\subsection{Weak Decay Form Factors}

Heavy-quark symmetry implies relations between the weak decay form
factors of heavy mesons, which are of particular interest. These
relations have been derived by Isgur and Wise~\cite{Isgu},
generalizing ideas developed by Nussinov and Wetzel~\cite{Nuss}, and
by Voloshin and Shifman~\cite{Vol1,Vol2}.

\begin{figure}[htb]
   \epsfxsize=7cm
   \centerline{\epsffile{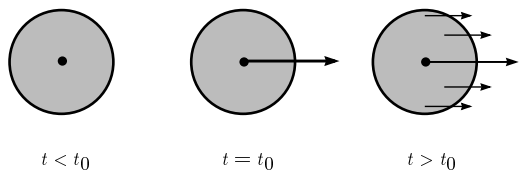}}
\caption{\label{fig:3.3}
Elastic transition induced by an external heavy-quark current.}
\end{figure}

Consider the elastic scattering of a $B$ meson, $\bar B(v)\to\bar
B(v')$, induced by a vector current coupled to the $b$ quark. Before
the action of the current, the light degrees of freedom inside the
$B$ meson orbit around the heavy quark, which acts as a static source
of colour. On average, the $b$ quark and the $B$ meson have the same
velocity $v$. The action of the current is to replace instantaneously
(at time $t=t_0$) the colour source by one moving at a velocity $v'$,
as indicated in Fig.~\ref{fig:3.3}. If $v=v'$, nothing happens; the
light degrees of freedom do not realize that there was a current
acting on the heavy quark. If the velocities are different, however,
the light constituents suddenly find themselves interacting with a
moving colour source. Soft gluons have to be exchanged to rearrange
them so as to form a $B$ meson moving at velocity $v'$. This
rearrangement leads to a form-factor suppression, reflecting the fact
that, as the velocities become more and more different, the
probability for an elastic transition decreases. The important
observation is that, in the limit $m_b\to\infty$, the form factor can
only depend on the Lorentz boost $\gamma = v\cdot v'$ connecting the
rest frames of the initial- and final-state mesons. Thus, in this
limit a dimensionless probability function $\xi(v\cdot v')$ describes
the transition. It is called the Isgur-Wise function~\cite{Isgu}. In
the HQET, which provides the appropriate framework for taking the
limit $m_b\to\infty$, the hadronic matrix element describing the
scattering process can thus be written as
\begin{equation}\label{elast}
   {1\over m_B}\,\langle\bar B(v')|\,\bar b_{v'}\gamma^\mu b_v\,
   |\bar B(v)\rangle = \xi(v\cdot v')\,(v+v')^\mu \,.
\end{equation}
Here $b_v$ and $b_{v'}$ are the velocity-dependent heavy-quark
fields of the HQET. It is important that the function $\xi(v\cdot
v')$ does not depend on $m_b$. The factor $1/m_B$ on the left-hand
side compensates for a trivial dependence on the heavy-meson mass
caused by the relativistic normalization of meson states, which is
conventionally taken to be
\begin{equation}\label{nonrelnorm}
   \langle\bar B(p')|\bar B(p)\rangle = 2 m_B v^0\,(2\pi)^3\,
   \delta^3(\vec p-\vec p\,') \,.
\end{equation}
Note that there is no term proportional to $(v-v')^\mu$ in
(\ref{elast}). This can be seen by contracting the matrix element
with $(v-v')_\mu$, which must give zero since $\rlap/v\,b_v = b_v$
and
$\bar b_{v'}\rlap/v' = \bar b_{v'}$.

It is more conventional to write the above matrix element in terms of
an elastic form factor $F_{\rm el}(q^2)$ depending on the momentum
transfer $q^2=(p-p')^2$:
\begin{equation}
   \langle\bar B(v')|\,\bar b\,\gamma^\mu b\,|\bar B(v)\rangle
   = F_{\rm el}(q^2)\,(p+p')^\mu \,,
\end{equation}
where $p^(\phantom{}'\phantom{}^)=m_B v^(\phantom{}'\phantom{}^)$.
Comparing this with (\ref{elast}), we find that
\begin{equation}
   F_{\rm el}(q^2) = \xi(v\cdot v') \,, \qquad
   q^2 = -2 m_B^2 (v\cdot v'-1) \,.
\end{equation}
Because of current conservation, the elastic form factor is
normalized to unity at $q^2=0$. This condition implies the
normalization of the Isgur-Wise function at the kinematic point
$v\cdot v'=1$, i.e.\ for $v=v'$:
\begin{equation}\label{Jcons2}
   \xi(1) = 1 \,.
\end{equation}
It is in accordance with the intuitive argument that the probability
for an elastic transition is unity if there is no velocity change.
Since for $v=v'$ the final-state meson is at rest in the rest frame
of
the initial meson, the point $v\cdot v'=1$ is referred to as the
zero-recoil limit.

The heavy-quark flavour symmetry can be used to replace the $b$ quark
in the final-state meson by a $c$ quark, thereby turning the $B$
meson into a $D$ meson. Then the scattering process turns into a weak
decay process. In the infinite-mass limit, the replacement $b_{v'}\to
c_{v'}$ is a symmetry transformation, under which the effective
Lagrangian is invariant. Hence, the matrix element
\begin{equation}
   {1\over\sqrt{m_B m_D}}\,\langle D(v')|\,\bar c_{v'}\gamma^\mu
   b_v\,|\bar B(v)\rangle = \xi(v\cdot v')\,(v+v')^\mu
\end{equation}
is still determined by the same function $\xi(v\cdot v')$. This is
interesting, since in general the matrix element of a
flavour-changing current between two pseudoscalar mesons is described
by two form factors:
\begin{equation}
   \langle D(v')|\,\bar c\,\gamma^\mu b\,|\bar B(v)\rangle
   = f_+(q^2)\,(p+p')^\mu - f_-(q^2)\,(p-p')^\mu \,.
\end{equation}
Comparing the above two equations, we find that
\begin{eqnarray}\label{inelast}
   f_\pm(q^2) &=& {m_B\pm m_D\over 2\sqrt{m_B m_D}}\,\xi(v\cdot v')
    \,, \nonumber\\
   q^2 &=& m_B^2 + m_D^2 - 2 m_B m_D\,v\cdot v' \,.
\end{eqnarray}
Thus, the heavy-quark flavour symmetry relates two a priori
independent form factors to one and the same function. Moreover, the
normalization of the Isgur-Wise function at $v\cdot v'=1$ now
implies a non-trivial normalization of the form factors $f_\pm(q^2)$
at the point of maximum momentum transfer, $q_{\rm max}^2=
(m_B-m_D)^2$:
\begin{equation}
   f_\pm(q_{\rm max}^2) = {m_B\pm m_D\over 2\sqrt{m_B m_D}} \,.
\end{equation}

The heavy-quark spin symmetry leads to additional relations among
weak decay form factors. It can be used to relate matrix elements
involving vector mesons to those involving pseudoscalar mesons. A
vector meson with longitudinal polarization is related to a
pseudoscalar meson by a rotation of the heavy-quark spin. Hence, the
spin-symmetry transformation $c_{v'}^\Uparrow\to c_{v'}^\Downarrow$
relates $\bar B\to D$ with $\bar B\to D^*$ transitions. The result of
this transformation is~\cite{Isgu}:
\begin{eqnarray}
   {1\over\sqrt{m_B m_{D^*}}}\,
   \langle D^*(v',\varepsilon)|\,\bar c_{v'}\gamma^\mu b_v\,
   |\bar B(v)\rangle &=& i\epsilon^{\mu\nu\alpha\beta}\,
    \varepsilon_\nu^*\,v'_\alpha v_\beta\,\,\xi(v\cdot v') \,,
    \nonumber\\
   {1\over\sqrt{m_B m_{D^*}}}\,
   \langle D^*(v',\varepsilon)|\,\bar c_{v'}\gamma^\mu\gamma_5\,
   b_v\,|\bar B(v)\rangle &=& \Big[ \varepsilon^{*\mu}\,(v\cdot v'+1)
    - v'^\mu\,\varepsilon^*\!\cdot v \Big] \xi(v\cdot v') \,,
    \nonumber\\
\end{eqnarray}
where $\varepsilon$ denotes the polarization vector of the $D^*$
meson. Once again, the matrix elements are completely described in
terms of the Isgur-Wise function. Now this is even more remarkable,
since in general four form factors, $V(q^2)$ for the vector current,
and $A_i(q^2)$, $i=0,1,2$, for the axial current, are required to
parametrize these matrix elements. In the heavy-quark limit, they
obey the relations~\cite{Neu1}
\begin{eqnarray}\label{PVff}
   {m_B+m_{D^*}\over 2\sqrt{m_B m_{D^*}}}\,\xi(v\cdot v')
   &=& V(q^2) = A_0(q^2) = A_1(q^2) \nonumber\\
   &=& \bigg[ 1 - {q^2\over(m_B+m_D)^2} \bigg]^{-1}\,A_1(q^2) \,,
    \nonumber\\
   \phantom{ \Bigg[ }
   q^2 &=& m_B^2 + m_{D^*}^2 - 2 m_B m_{D^*}\,v\cdot v' \,.
\end{eqnarray}

Equations (\ref{inelast}) and (\ref{PVff}) summarize the relations
imposed by heavy-quark symmetry on the weak decay form factors
describing the semileptonic decay processes $\bar B\to
D\,\ell\,\bar\nu$ and $\bar B\to D^*\ell\,\bar\nu$. These relations
are model-independent consequences of QCD in the limit where $m_b,
m_c\gg\Lambda_{\rm QCD}$. They play a crucial role in the
determination of the CKM matrix element $|V_{cb}|$. In terms of the
recoil variable $w=v\cdot v'$, the differential semileptonic decay
rates in the heavy-quark limit become~\cite{Vcb}:
\begin{eqnarray}\label{rates}
   {{\rm d}\Gamma(\bar B\to D\,\ell\,\bar\nu)\over{\rm d}w}
   &=& {G_F^2\over 48\pi^3}\,|V_{cb}|^2\,(m_B+m_D)^2\,m_D^3\,
    (w^2-1)^{3/2}\,\xi^2(w) \,, \nonumber\\
   {{\rm d}\Gamma(\bar B\to D^*\ell\,\bar\nu)\over{\rm d}w}
   &=& {G_F^2\over 48\pi^3}\,|V_{cb}|^2\,(m_B-m_{D^*})^2\,
    m_{D^*}^3\,\sqrt{w^2-1}\,(w+1)^2 \nonumber\\
   &&\times \Bigg[ 1 + {4w\over w+1}\,
    {m_B^2 - 2 w\,m_B m_{D^*} + m_{D^*}^2\over(m_B-m_{D^*})^2}
    \Bigg]\,\xi^2(w) \,.
\end{eqnarray}
These expressions receive symmetry-breaking corrections, since the
masses of the heavy quarks are not infinitely large. Perturbative
corrections of order $\alpha_s^n(m_Q)$ can be calculated order by
order in perturbation theory. A more difficult task is to control the
non-perturbative power corrections of order $(\Lambda_{\rm
QCD}/m_Q)^n$. The HQET provides a systematic framework for analysing
these corrections. For the case of weak-decay form factors, the
analysis of the $1/m_Q$ corrections was performed by
Luke~\cite{Luke}. Later, Falk and the present author have analysed
the structure of $1/m_Q^2$ corrections for both meson and baryon weak
decay form factors~\cite{FaNe}. We shall not discuss these rather
technical issues in detail, but only mention the most important
result of Luke's analysis. It concerns the zero-recoil limit, where
an analogue of the Ademollo-Gatto theorem~\cite{AGTh} can be proved.
This is Luke's theorem~\cite{Luke}, which states that the matrix
elements describing the leading $1/m_Q$ corrections to weak decay
amplitudes vanish at zero recoil. This theorem is valid to all orders
in perturbation theory~\cite{FaNe,Neu7,ChGr}. Most importantly, it
protects the $\bar B\to D^*\ell\,\bar\nu$ decay rate from receiving
first-order $1/m_Q$ corrections at zero recoil~\cite{Vcb}. [A similar
statement is not true for the decay $\bar B\to D\,\ell\,\bar\nu$. The
reason is simple but somewhat subtle. Luke's theorem protects only
those form factors not multiplied by kinematic factors that vanish
for $v=v'$. By angular momentum conservation, the two pseudoscalar
mesons in the decay $\bar B\to D\,\ell\,\bar\nu$ must be in a
relative $p$ wave, and hence the amplitude is proportional to the
velocity $|\vec v_D|$ of the $D$ meson in the $B$-meson rest frame.
This leads to a factor $(w^2-1)$ in the decay rate. In such a
situation, kinematically suppressed form factors can
contribute~\cite{Neu1}.]

\subsection{Short-Distance Corrections}

In section~\ref{sec:2}, we have discussed the first two steps in the
construction of the HQET. Integrating out the small components in the
heavy-quark fields, a non-local effective action was derived, which
was then expanded in a series of local operators. The effective
Lagrangian obtained that way correctly reproduces the long-distance
physics of the full theory (see Fig.~\ref{fig:magic}). It does not
contain the short-distance physics correctly, however. The reason is
obvious: a heavy quark participates in strong interactions through
its coupling to gluons. These gluons can be soft or hard, i.e.\ their
virtual momenta can be small, of the order of the confinement scale,
or large, of the order of the heavy-quark mass. But hard gluons can
resolve the spin and flavour quantum numbers of a heavy quark. Their
effects lead to a renormalization of the coefficients of the
operators in the HQET. A new feature of such short-distance
corrections is that through the running coupling constant they induce
a logarithmic dependence on the heavy-quark mass~\cite{Vol1}. Since
$\alpha_s(m_Q)$ is small, these effects can be calculated in
perturbation theory.

Consider, as an example, the matrix elements of the vector current
$V=\bar q\,\gamma^\mu Q$. In QCD this current is partially conserved
and needs no renormalization~\cite{Prep}. Its matrix elements are
free of ultraviolet divergences. Still, these matrix elements have a
logarithmic dependence on $m_Q$ from the exchange of hard gluons with
virtual momenta of the order of the heavy-quark mass. If one goes
over to the effective theory by taking the limit $m_Q\to\infty$,
these logarithms diverge. Consequently, the vector current in the
effective theory does require a renormalization~\cite{PoWi}. Its
matrix elements depend on an arbitrary renormalization scale $\mu$,
which separates the regions of short- and long-distance physics. If
$\mu$ is chosen such that $\Lambda_{\rm QCD}\ll\mu\ll m_Q$, the
effective coupling constant in the region between $\mu$ and $m_Q$ is
small, and perturbation theory can be used to compute the
short-distance corrections. These corrections have to be added to the
matrix elements of the effective theory, which contain the
long-distance physics below the scale $\mu$. Schematically, then, the
relation between matrix elements in the full and in the effective
theory is
\begin{equation}\label{OPEex}
   \langle\,V(m_Q)\,\rangle_{\rm QCD}
   = C_0(m_Q,\mu)\,\langle V_0(\mu)\rangle_{\rm HQET}
   + {C_1(m_Q,\mu)\over m_Q}\,\langle V_1(\mu)\rangle_{\rm HQET}
   + \ldots \,,
\end{equation}
where we have indicated that matrix elements in the full theory
depend on $m_Q$, whereas matrix elements in the effective theory are
mass-independent, but do depend on the renormalization scale. The
Wilson coefficients $C_i(m_Q,\mu)$ are defined by this relation.
Order by order in perturbation theory, they can be computed from a
comparison of the matrix elements in the two theories. Since the
effective theory is constructed to reproduce correctly the low-energy
behaviour of the full theory, this ``matching'' procedure is
independent of any long-distance physics, such as infrared
singularities, non-perturbative effects, the nature of the external
states used in the matrix elements, etc.

The calculation of the coefficient functions in perturbation theory
uses the powerful methods of the renormalization group. It is in
principle straightforward, yet in practice rather tedious. A
comprehensive discussion of most of the existing calculations of
short-distance corrections in the HQET can be found in
Ref.~24.

\subsection{Model-Independent Determination of $|V_{cb}|$}

We will now discuss some of the most important applications and tests
of the above formalism in the context of semileptonic decays of $B$
mesons. A model-independent determination of the CKM matrix element
$|V_{cb}|$ based on heavy-quark symmetry can be obtained by measuring
the recoil spectrum of $D^*$ mesons produced in $\bar B\to
D^*\ell\,\bar\nu$ decays~\cite{Vcb}. In the heavy-quark limit, the
differential decay rate for this process has been given in
(\ref{rates}). In order to allow for corrections to that limit, we
write
\begin{eqnarray}
   {{\rm d}\Gamma(\bar B\to D^*\ell\,\bar\nu)\over{\rm d}w}
   &=& {G_F^2\over 48\pi^3}\,(m_B-m_{D^*})^2\,m_{D^*}^3
    \sqrt{w^2-1}\,(w+1)^2 \nonumber\\
   &&\mbox{}\times \Bigg[ 1 + {4w\over w+1}\,
    {m_B^2-2w\,m_B m_{D^*} + m_{D^*}^2\over(m_B - m_{D^*})^2}
    \Bigg]\,|V_{cb}|^2\,{\cal{F}}^2(w) \,, \nonumber\\
\end{eqnarray}
where the hadronic form factor ${\cal F}(w)$ coincides with the
Isgur-Wise function up to symmetry-breaking corrections of order
$\alpha_s(m_Q)$ and $\Lambda_{\rm QCD}/m_Q$. The idea is to measure
the product $|V_{cb}|\,{\cal F}(w)$ as a function of $w$, and to
extract $|V_{cb}|$ from an extrapolation of the data to the
zero-recoil point $w=1$, where the $B$ and the $D^*$ mesons have a
common rest frame. At this kinematic point, heavy-quark symmetry
helps us to calculate the normalization ${\cal F}(1)$ with small and
controlled theoretical errors. Since the range of $w$ values
accessible in this decay is rather small ($1<w<1.5$), the
extrapolation can be done using an expansion around $w=1$:
\begin{equation}\label{Fexp}
   {\cal F}(w) = {\cal F}(1)\,\Big[ 1 - \widehat\varrho^2\,(w-1)
   + \widehat c\,(w-1)^2 \dots \Big] \,.
\end{equation}
The slope $\widehat\varrho^2$ and the curvature $\widehat c$ are
treated as parameters.

\begin{figure}[htb]
   \epsfxsize=8cm
   \vspace{0.3cm}
   \centerline{\epsffile{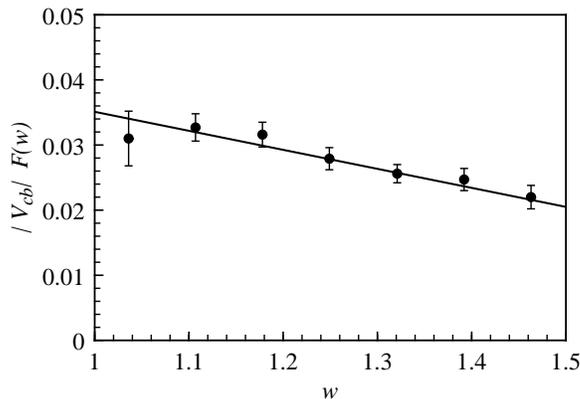}}
   \vspace{-0.3cm}
\caption{\label{fig:CLVcb}
CLEO data for the product $|V_{cb}|\,{\cal F}(w)$, as extracted from
the recoil spectrum in $\bar B\to D^*\ell\,\bar\nu$
decays~\protect\cite{CLEOVcb}. The line shows a linear fit to the
data.}
\end{figure}

Measurements of the recoil spectrum have been performed first by the
AR\-GUS~\cite{ARGVcb} and CLEO~\cite{CLEOVcb} Collaborations in
experiments operating at the $\Upsilon(4s)$ resonance, and more
recently by the ALEPH~\cite{ALEVcb}, DELPHI~\cite{DELVcb} and
OPAL~\cite{OPALVcb} Collaborations at LEP. As an example,
Fig.~\ref{fig:CLVcb} shows the data reported by the CLEO
Collaboration. The results obtained by the various experimental
groups are summarized in Table~\ref{tab:Vcb}. In the first analyses,
the curvature term in (\ref{Fexp}) was omitted, and the data were
fitted with a linear form factor. Later, the effect of a non-zero
curvature has been taken into account~\cite{ALEVcb,OPALVcb}. It can
be shown in a model-independent way that the shape of the form factor
is highly constrained by analyticity and unitarity
requirements~\cite{Boyd2,Capr}. In particular, the curvature at $w=1$
is strongly correlated with the slope of the form factor. For the
values of $\widehat\varrho^2$ given in Table~\ref{tab:Vcb}, one
obtains a small positive curvature, so that the results obtained from
a linear fit are very little affected by including a curvature term.
The weighted average of the experimental results is
\begin{eqnarray}\label{VcbF}
   |V_{cb}|\,{\cal F}(1) &=& (34.1\pm 1.4)\times 10^{-3} \,,
    \nonumber\\
   \widehat\varrho^2 &=& 0.80\pm 0.09 \,.
\end{eqnarray}

\begin{table}[htb]
\caption{\label{tab:Vcb}
Values for $|V_{cb}|\,{\cal F}(1)$ (in units of $10^{-3}$) and
$\widehat\varrho^2$ extracted from measurements of the recoil
spectrum in $\bar B\to D^*\ell\,\bar\nu$ decays.}
\vspace{0.5cm}
\centerline{\begin{tabular}{|l|l|l|l|}\hline\hline
Reference \rule[-0.25cm]{0cm}{0.7cm} & Method & $|V_{cb}|\,
 {\cal F}(1)~(10^{-3})$ & \hspace{11mm} $\widehat\varrho^2$ \\
\hline
ARGUS~\cite{ARGVcb} \rule{0cm}{0.4cm} & Linear Fit &
 $38.8\pm 4.3\pm 2.5$ & $1.17\pm 0.22\pm 0.06$ \\
CLEO~\cite{CLEOVcb} & Linear Fit & $35.1\pm 1.9\pm 2.0$ &
 $0.84\pm 0.12\pm 0.08$ \\
ALEPH~\cite{ALEVcb} & Quadratic Fit & $32.0\pm 2.1\pm 2.0$ &
 $0.37\pm 0.26\pm 0.14$ \\
 & Linear Fit & $31.9\pm 1.8\pm 1.9$ & $0.31\pm 0.17\pm 0.08$ \\
DELPHI~\cite{DELVcb} & Linear Fit & $35.0\pm 1.9\pm 2.3$ &
 $0.81\pm 0.16\pm 0.10$ \\
OPAL~\cite{OPALVcb} & Quadratic Fit & $32.8\pm 1.9\pm 2.2$ &
 $0.55\pm 0.24\pm 0.05$ \\
 & Linear Fit & $32.5\pm 1.7$ & $0.42\pm 0.17$ \\[0.06cm]
\hline\hline
\end{tabular}}
\end{table}

Heavy-quark symmetry implies that the general structure of the
symmetry-breaking corrections to the form factor at zero recoil
is~\cite{Vcb}
\begin{equation}
   {\cal F}(1) = \eta_A\,\bigg( 1 + 0\times
   {\Lambda_{\rm QCD}\over m_Q}
   + \mbox{const}\times {\Lambda_{\rm QCD}^2\over m_Q^2}
   + \dots \bigg) \equiv \eta_A\,(1+\delta_{1/m^2}) \,,
\end{equation}
where $\eta_A$ is a short-distance correction arising from the finite
renormalization of the flavour-changing axial current at zero recoil,
and $\delta_{1/m^2}$ parametrizes second-order (and higher) power
corrections. The absence of first-order power corrections at zero
recoil is a consequence of Luke's theorem~\cite{Luke}. The one-loop
expression for $\eta_A$ has been known for a long
time~\cite{Pasc,Vol2,QCD1}:
\begin{equation}\label{etaA1}
   \eta_A = 1 + {\alpha_s(M)\over\pi}\,\bigg(
   {m_b+m_c\over m_b-m_c}\,\ln{m_b\over m_c} - {8\over 3} \bigg)
   \approx 0.96 \,.
\end{equation}
The scale $M$ in the running coupling constant can be fixed by
adopting the prescription of Brodsky, Lepage and Mackenzie
(BLM)~\cite{BLM}, where it is identified with the average virtuality
of the gluon in the one-loop diagrams that contribute to $\eta_A$. If
$\alpha_s(M)$ is defined in the $\overline{\mbox{\sc ms}}$ scheme,
the result is~\cite{etaVA} $M\approx 0.51\sqrt{m_c m_b}$. Several
estimates of higher-order corrections to $\eta_A$ have been
discussed. A renormalization-group resummation of logarithms of the
type $(\alpha_s\ln m_b/m_c)^n$, $\alpha_s(\alpha_s\ln m_b/m_c)^n$ and
$m_c/m_b(\alpha_s\ln m_b/m_c)^n$ leads
to~\cite{PoWi,JiMu}$^-$\cite{QCD2} $\eta_A\approx 0.985$. On the
other
hand, a resummation of ``renormalon-chain'' contributions of the form
$\beta_0^{n-1}\alpha_s^n$, where $\beta_0=11-\frac{2}{3}n_f$ is the
first coefficient of the QCD $\beta$-function, gives~\cite{flow}
$\eta_A\approx 0.945$. Using these partial resummations to estimate
the uncertainty gives $\eta_A = 0.965\pm 0.020$. Recently, Czarnecki
has improved this estimate by calculating $\eta_A$ at two-loop
order~\cite{Czar}. His result,
\begin{equation}
   \eta_A = 0.960\pm 0.007 \,,
\end{equation}
is in excellent agreement with the BLM-improved one-loop expression
(\ref{etaA1}). Here the error is taken to be the size of the two-loop
correction.

The analysis of the power corrections is more difficult, since it
cannot rely on perturbation theory. Three approaches have been
discussed: in the ``exclusive approach'', all $1/m_Q^2$ operators in
the HQET are classified and their matrix elements estimated, leading
to~\cite{FaNe,TMann} $\delta_{1/m^2}=-(3\pm 2)\%$; the ``inclusive
approach'' has been used to derive the bound $\delta_{1/m^2}<-3\%$,
and to estimate that~\cite{Shif} $\delta_{1/m^2}=-(7\pm 3)\%$; the
``hybrid approach'' combines the virtues of the former two to obtain
a more restrictive lower bound on $\delta_{1/m^2}$. This leads
to~\cite{Vcbnew}
\begin{equation}
   \delta_{1/m^2} = - 0.055\pm 0.025 \,.
\end{equation}

Combining the above results, adding the theoretical errors linearly
to be conservative, gives
\begin{equation}\label{F1}
   {\cal F}(1) = 0.91\pm 0.03
\end{equation}
for the normalization of the hadronic form factor at zero recoil.
Thus, the corrections to the heavy-quark limit amount to a moderate
decrease of the form factor of about 10\%. This can be used to
extract from the experimental result (\ref{VcbF}) the
model-independent value
\begin{equation}\label{Vcbexc}
   |V_{cb}| = (37.5\pm 1.5_{\rm exp}\pm 1.2_{\rm th})
   \times 10^{-3} \,.
\end{equation}

\subsection{Bounds and Predictions for $\widehat\varrho^2$}

The slope parameter $\widehat\varrho^2$ in the expansion of the
physical form factor in (\ref{Fexp}) differs from the slope parameter
$\varrho^2$ of the Isgur-Wise function by corrections that violate
the heavy-quark symmetry. The short-distance corrections have been
calculated, with the result~\cite{Vcbnew}
\begin{equation}\label{rhorel}
   \widehat\varrho^2 = \varrho^2 + (0.16\pm 0.02) + O(1/m_Q) \,.
\end{equation}
Bjorken has shown that the slope of the Isgur-Wise function is
related to the form factors of transitions of a ground-state heavy
meson into excited states, in which the light degrees of freedom
carry quantum numbers $j^P=\frac{1}{2}^+$ or $\frac{3}{2}^+$, by a
sum rule which is an expression of quark-hadron duality: in the
heavy-quark limit, the inclusive sum of the probabilities for
decays into hadronic states is equal to the probability for the free
quark transition. If one normalizes the latter probability to unity,
the sum rule takes the form~\cite{Bjor}$^-$\cite{BjDT}
\begin{eqnarray}\label{inclsum}
   1 &=& {w+1\over 2}\,\bigg\{ |\xi(w)|^2 + \sum_l |\xi^{(l)}(w)|^2
    \bigg\} \nonumber\\
   &&\mbox{}+ (w-1)\,\bigg\{ 2\sum_m |\tau_{1/2}^{(m)}(w)|^2
    + (w+1)^2 \sum_n |\tau_{3/2}^{(n)}(w)|^2 \bigg\}
    + O\big[(w-1)^2\big] \,, \nonumber\\
\end{eqnarray}
where $l,m,n$ label the radial excitations of states with the same
spin-parity quantum numbers. The terms in the first line on the
right-hand side of the sum rule correspond to transitions into states
containing light constituents with quantum numbers
$j^P=\frac{1}{2}^-$. The ground state gives a contribution
proportional to the Isgur-Wise function, and excited states
contribute proportionally to analogous functions $\xi^{(l)}(w)$.
Because at zero recoil these states must be orthogonal to the ground
state, it follows that $\xi^{(l)}(1)=0$, and the corresponding
contributions to (\ref{inclsum}) are of order $(w-1)^2$. The
contributions in the second line correspond to transitions into
states with $j^P=\frac{1}{2}^+$ or $\frac{3}{2}^+$. Because of the
change in parity, these are $p$-wave transitions. The amplitudes are
proportional to the velocity $|\vec v_f|= (w^2-1)^{1/2}$ of the final
state in the rest frame of the initial state, which explains the
suppression factor $(w-1)$ in the decay probabilities. The functions
$\tau_j(w)$ are the analogues of the Isgur-Wise function for these
transitions~\cite{IsgW}. Transitions into excited states with quantum
numbers other than the above proceed via higher partial waves and are
suppressed by at least a factor $(w-1)^2$.

For $w=1$, eq.~(\ref{inclsum}) reduces to the normalization condition
for the Isgur-Wise function. The Bjorken sum rule is obtained by
expanding in powers of $(w-1)$ and keeping terms of first order.
Taking into account the definition of the slope parameter,
$\xi'(1)=-\varrho^2$, one finds that~\cite{Bjor,IsgW}
\begin{equation}\label{Bjsr}
   \varrho^2 = {1\over 4} + \sum_m |\tau_{1/2}^{(m)}(1)|^2
   + 2 \sum_n |\tau_{3/2}^{(n)}(1)|^2 > {1\over 4} \,.
\end{equation}
Notice that the lower bound is due to the prefactor
$\frac{1}{2}(w+1)$ of the first term in (\ref{inclsum}) and is of
purely kinematic origin. In the analogous sum rule for
$\Lambda_Q$ baryons, this factor is absent, and consequently the
slope parameter of the baryon Isgur-Wise function is only subject to
the trivial constraint~\cite{Neu1,IWYo} $\varrho^2>0$.

Voloshin has derived another sum rule involving the form factors for
transitions into excited states, which is the analogue of the
``optical sum rule'' for the dipole scattering of light in atomic
physics. It reads~\cite{Volsum}
\begin{equation}
   {m_M-m_Q\over 2} = \sum_m E_{1/2}^{(m)}\,|\tau_{1/2}^{(m)}(1)|^2
   + 2 \sum_n  E_{3/2}^{(n)}\,|\tau_{3/2}^{(n)}(1)|^2 \,,
\end{equation}
where $E_j$ are the excitation energies relative to the mass $m_M$ of
the ground-state heavy meson. This relation can be combined with the
Bjorken sum rule to obtain an upper bound for the slope parameter
$\varrho^2$:
\begin{equation}\label{Volsr}
   \varrho^2 < {1\over 4} + {m_M-m_Q\over 2 E_{\rm min}} \,,
\end{equation}
where $E_{\rm min}$ denotes the minimum excitation energy. In the
quark model, one expects that $E_{\rm min}\approx m_M-m_Q$, and one
may use this as an estimate to obtain
$\varrho^2<0.75$.\footnote{Strictly speaking, the lowest excited
state contributing to the sum rule is $D+\pi$, which has an
excitation-energy spectrum with a threshold at $m_\pi$. However, this
spectrum is broad, so that this contribution will not invalidate the
upper bound for $\varrho^2$ derived here.}

\begin{figure}[htb]
   \epsfxsize=8cm
   \centerline{\epsffile{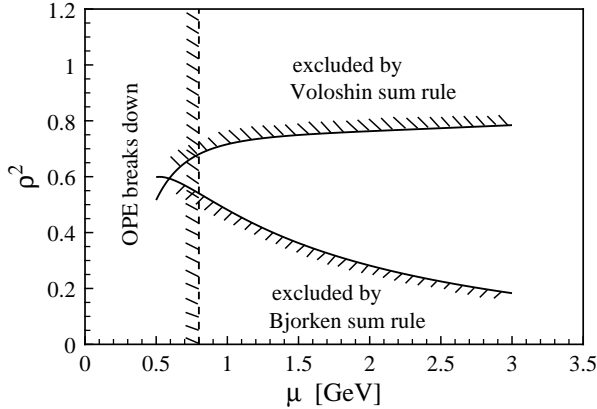}}
\caption{\label{fig:2}
Bounds for the slope parameter $\varrho^2$ following from
the Bjorken and Voloshin sum rules.}
\end{figure}

The above discussion of the sum rules ignores renormalization
effects. Both perturbative and non-perturbative corrections to
(\ref{Bjsr}) and (\ref{Volsr}) can be incorporated using the OPE,
where one introduces a momentum scale $\mu\sim\mbox{few}\times
\Lambda_{\rm QCD}$ large enough for $\alpha_s(\mu)$ and power
corrections of order $(\Lambda_{\rm QCD}/\mu)^n$ to be small, but
otherwise as small as possible so as to suppress the contributions
from excited states~\cite{GrKo,BLRW}. The result is~\cite{KoNe}
$\varrho_{\rm min}^2(\mu) < \varrho^2 < \varrho_{\rm max}^2(\mu)$,
where the boundary values are shown in Fig.~\ref{fig:2} as a function
of the scale $\mu$. Assuming that the OPE works down to values
$\mu\approx 0.8$~GeV, one obtains rather tight bounds for the slope
parameters:
\begin{equation}\label{rhobounds}
   0.5 < \varrho^2 < 0.8 \,, \qquad
   0.5 < \widehat\varrho^2 < 1.1 \,.
\end{equation}
The allowed region for $\widehat\varrho^2$ has been increased in
order to account for the unknown $1/m_Q$ corrections in the relation
(\ref{rhorel}). The experimental result given in (\ref{VcbF}) falls
inside this region.

These bounds compare well with theoretical predictions for the slope
parameters. QCD sum rules have been used to calculate the slope of
the Isgur-Wise function. The results obtained by various authors are
$\varrho^2 = 0.84\pm 0.02$ (Bagan et al.~\cite{Baga}), $0.7\pm 0.1$
(present author~\cite{twoloop}), $0.70\pm 0.25$ (Blok and
Shifman~\cite{BlSh}), and $1.00\pm 0.02$ (Narison~\cite{Nari}). The
UKQCD Collaboration has presented a lattice calculation of the slope
of the form factor ${\cal F}(w)$, yielding~\cite{Lattrho}
$\widehat\varrho^2 = 0.9_{-0.3-0.2}^{+0.2+0.4}$. We stress that the
sum-rule bounds in (\ref{rhobounds}) are largely model independent;
model calculations in strong disagreement with these bounds should be
discarded.

\subsection{Measurements of $\bar B\to D^*\ell\,\bar\nu$ and $\bar
B\to D\,\ell\,\bar\nu$ Form Factors and Tests of Heavy-Quark
Symmetry}

We have discussed earlier in this section that heavy-quark symmetry
implies relations between the semileptonic form factors of heavy
mesons. They receive symmetry-breaking corrections, which can be
estimated using the HQET. The extent to which these relations hold
can be tested experimentally by comparing the different form factors
describing the decays $\bar B\to D^{(*)}\ell\,\bar\nu$ at the same
value of $w$.

When the lepton mass is neglected, the differential decay
distributions in $\bar B\to D^*\ell\,\bar\nu$ decays can be
parametrized by three helicity amplitudes, or equivalently by three
independent combinations of form factors. It has been suggested that
a good choice for three such quantities should be inspired by the
heavy-quark limit~\cite{review,subl}. One thus defines a form factor
$h_{A1}(w)$, which up to symmetry-breaking corrections coincides with
the Isgur-Wise function, and two form-factor ratios
\begin{eqnarray}
   R_1(w) &=& \bigg[ 1 - {q^2\over(m_B+m_{D^*})^2} \bigg]\,
    {V(q^2)\over A_1(q^2)} \,, \nonumber\\
   R_2(w) &=& \bigg[ 1 - {q^2\over(m_B+m_{D^*})^2} \bigg]\,
    {A_2(q^2)\over A_1(q^2)} \,.
\end{eqnarray}
The relation between $w$ and $q^2$ has been given in (\ref{PVff}).
This definition is such that in the heavy-quark limit
$R_1(w)=R_2(w)=1$ independently of $w$.

To extract the functions $h_{A1}(w)$, $R_1(w)$ and $R_2(w)$ from
experimental data is a complicated task. However, HQET-based
calculations suggest that the $w$ dependence of the form-factor
ratios, which is induced by symmetry-breaking effects, is rather
mild~\cite{subl}. Moreover, the form factor $h_{A1}(w)$ is expected
to
have a nearly linear shape over the accessible $w$ range. This
motivates to introduce three parameters $\varrho_{A1}^2$, $R_1$ and
$R_2$ by
\begin{eqnarray}
   h_{A1}(w) &\approx& {\cal F}(1)\,\Big[ 1 - \varrho_{A1}^2 (w-1)
    \Big] \,, \nonumber\\
   R_1(w) &\approx& R_1 \,, \qquad
   R_2(w) \approx R_2 \,,
\end{eqnarray}
where ${\cal F}(1)=0.91\pm 0.03$ from (\ref{F1}). The CLEO
Collaboration has extracted these three parameters from an analysis
of the angular distributions in $\bar B\to D^*\ell\,\bar\nu$
decays~\cite{CLEff}. The results are
\begin{equation}
   \varrho_{A1}^2 = 0.91\pm 0.16 \,, \qquad
   R_1 = 1.18\pm 0.32 \,, \qquad
   R_2 = 0.71\pm 0.23 \,.
\end{equation}
Using the HQET, one obtains an essentially model-independent
prediction for the symme\-try-breaking corrections to $R_1$, whereas
the corrections to $R_2$ are somewhat model dependent. To good
approximation~\cite{review}
\begin{eqnarray}
   R_1 &\approx& 1 + {4\alpha_s(m_c)\over 3\pi}
    + {\bar\Lambda\over 2 m_c}\approx 1.3\pm 0.1 \,, \nonumber\\
   R_2 &\approx& 1 - \kappa\,{\bar\Lambda\over 2 m_c}
    \approx 0.8\pm 0.2 \,,
\end{eqnarray}
with $\kappa\approx 1$ from QCD sum rules~\cite{subl}. Here
$\bar\Lambda$ is the ``binding energy'' as defined in (\ref{Lbdef}).
Theoretical calculations~\cite{Lamsr1,Lamsr2} as well as
phenomenological analyses~\cite{GKLW,FLS2} suggest that
$\bar\Lambda\approx 0.45$--0.65~GeV is the appropriate value to be
used in one-loop calculations. A quark-model calculation of $R_1$ and
$R_2$ gives results similar to the HQET predictions~\cite{ClWa}:
$R_1\approx 1.15$ and $R_2\approx 0.91$. The experimental data
confirm the theoretical prediction that $R_1>1$ and $R_2<1$, although
the errors are still large.

There is a model-independent relation between the three parameters
determined from the analysis of angular distributions and the slope
parameter $\widehat\varrho^2$ extracted from the semileptonic
spectrum. It reads~\cite{Vcbnew}
\begin{equation}
   \varrho_{A1}^2 - \widehat\varrho^2 = {1\over 6}\,(R_1^2-1)
   + {m_B\over 3(m_B-m_{D^*})}\,(1-R_2) \,.
\end{equation}
The CLEO data give $0.07\pm 0.20$ for the difference of the slope
parameters on the left-hand side, and $0.22\pm 0.18$ for the
right-hand side. Both values are compatible within errors.

More recently, heavy-quark symmetry has also been tested by comparing
the form factor ${\cal F}(w)$ in $\bar B\to D^*\ell\,\bar\nu$ decays
with the corresponding form factor ${\cal G}(w)$ governing $\bar B\to
D\,\ell\,\bar\nu$ decays. The theoretical prediction~\cite{Capr,subl}
\begin{equation}
   \frac{{\cal G}(1)}{{\cal F}(1)} = 1.08\pm 0.06
\end{equation}
compares well with the experimental results for this ratio: $0.99\pm
0.19$ reported by the CLEO Collaboration~\cite{CLEOBD}, and $0.87\pm
0.30$ reported by the ALEPH Collaboration~\cite{ALEVcb}. In these
analyses, it has also been tested that within experimental errors the
shape of the two form factors agrees over the entire range of $w$
values.

The results of the analyses described above are very encouraging.
Within errors, the experiments confirm the HQET predictions, starting
to test them at the level of symmetry-breaking corrections.

\subsection{Decays to Charmless Final States}

For completeness, we will discuss briefly semileptonic $B$-meson
decays into charmless final states, although heavy-quark symmetry
does not help much in the analysis of these processes. Recently, the
CLEO Collaboration has reported a first signal for the exclusive
semileptonic decay modes $\bar B\to\pi\,\ell\,\bar\nu$ and $\bar
B\to\rho\,\ell\,\bar\nu$. The underlying quark process for these
transitions is $b\to u\,\ell\,\bar\nu$. Thus, these decays provide
information on the strength of the CKM matrix element $V_{ub}$. The
observed branching fractions are~\cite{Bpirho}:
\begin{eqnarray}\label{CLEOVub}
   \mbox{B}(\bar B\to\pi\,\ell\,\bar\nu)
   &=& (1.8\pm 0.5)\times 10^{-4} \,, \nonumber\\
   \mbox{B}(\bar B\to\rho\,\ell\,\bar\nu)
   &=& (2.5_{-0.9}^{+0.8})\times 10^{-4} \,.
\end{eqnarray}

\begin{table}[htb]
\caption{\label{tab:Vub}
Values for $|V_{ub}/V_{cb}|$ extracted from the CLEO measurement of
exclusive semileptonic $B$ decays into charmless final states, taking
$|V_{cb}|=0.040$. The first error quoted is experimental, the second
(when available) is theoretical.}
\vspace{0.5cm}
\centerline{\begin{tabular}{|l|l|c|c|}\hline\hline
Method \rule[-0.25cm]{0cm}{0.7cm} & Reference
 & $\bar B\to\pi\,\ell\,\bar\nu$ & $\bar B\to\rho\,\ell\,\bar\nu$ \\
\hline
Sum Rules \rule{0cm}{0.4cm} & Narison~\cite{SNar} &
 $0.177\pm 0.025\pm 0.001$ & $0.064_{-0.012}^{+0.010}\pm 0.003$ \\
 & Ball~\cite{PBal} & $0.117\pm 0.016\pm 0.012$ &
 $0.090_{-0.016}^{+0.014}\pm 0.015$ \\
 & Khod., R\"uckl~\cite{Khod} \hspace{-3mm} & $0.095\pm 0.013$ &
 --- \\[0.06cm]
\hline
Lattice QCD \rule{0cm}{0.4cm} & UKQCD~\cite{UKQCDVub} &
 $0.115\pm 0.016_{-0.011}^{+0.013}$ & --- \\
 & APE~\cite{APEVub} & $0.094\pm 0.013\pm 0.023$ & --- \\[0.06cm]
\hline
Pert.\ QCD \rule{0cm}{0.4cm} & Li, Yu~\cite{LiYu} &
 $0.060\pm 0.008$ & --- \\[0.06cm]
\hline
Quark Models \hspace{-3mm} \rule{0cm}{0.4cm} & BSW~\cite{BSW} &
 $0.098\pm 0.014$ & $0.061_{-0.011}^{+0.010}$ \\
 & KS~\cite{KS} & $0.098\pm 0.014$ & $0.054_{-0.007}^{+0.006}$ \\
 & ISGW2~\cite{ISGW2} & $0.086\pm 0.012$ &
 $0.083_{-0.015}^{+0.013}$ \\
 & Melikhov~\cite{Meli} & $0.099\pm 0.014\pm 0.014$ &
 $0.097_{-0.017}^{+0.015}\pm 0.014$ \\[0.06cm]
\hline\hline
\end{tabular}}
\end{table}

The theoretical description of these heavy-to-light ($b\to u$) decays
is more model dependent than that for heavy-to-heavy ($b\to c$)
transitions, because heavy-quark symmetry does not help to constrain
the relevant hadronic form factors. A variety of calculations for
such form factors exists, based on QCD sum rules, lattice gauge
theory, perturbative QCD, or quark models. Table~\ref{tab:Vub}
contains a summary of values extracted for the ratio
$|V_{ub}/V_{cb}|$ from a selection of such calculations. Some
approaches are more consistent than others in that the extracted
values are compatible for the two decay modes. With few exceptions,
the results lie in the range
\begin{equation}
   \left| {V_{ub}\over V_{cb}} \right|_{\rm excl}
   = 0.06\mbox{--}0.11 \,,
\end{equation}
which is in good agreement with the measurement of $|V_{ub}|$
obtained from the endpoint region of the lepton spectrum in inclusive
semileptonic decays~\cite{btou1,btou2}:
\begin{equation}\label{Vubval}
   \left| {V_{ub}\over V_{cb}} \right|_{\rm incl}
   = 0.08\pm 0.01_{\rm exp}\pm 0.02_{\rm th} \,.
\end{equation}
Clearly, this is only the first step towards a more reliable
determination of $|V_{ub}|$; yet, with the discovery of exclusive
$b\to u$ transitions an important milestone has been met. Efforts
must now concentrate on more reliable methods to determine the form
factors for heavy-to-light transitions. Some new ideas in this
direction have been discussed recently. They are based on lattice
calculations~\cite{Flynn}, analyticity constraints~\cite{Boyd1,Lell},
or variants of the form-factor relations for heavy-to-heavy
transitions~\cite{Stech}.

\section{Inclusive Decay Rates and Lifetimes}
\label{sec:4}

Inclusive decay rates determine the probability of the decay of a
particle into the sum of all possible final states with a given set
of global quantum numbers. An example is provided by the inclusive
semileptonic decay rate of the $B$ meson, $\Gamma(\bar B\to
X\,\ell\,\bar\nu)$, where the final state consists of a
lepton-neutrino pair accompanied by any number of hadrons. Here we
shall discuss the theoretical description of inclusive decays of
hadrons containing a heavy quark~\cite{Chay}$^-$\cite{MNTM}. From the
theoretical point of view, such decays have two advantages: first,
bound-state effects related to the initial state (such as the ``Fermi
motion'' of the heavy quark inside the hadron~\cite{shape,Fermi}) can
be accounted for in a systematic way using the heavy-quark expansion;
secondly, the fact that the final state consists of a sum over many
hadronic channels eliminates bound-state effects related to the
properties of individual hadrons. This second feature is based on the
hypothesis of quark-hadron duality, which is an important concept in
QCD phenomenology. The assumption of duality is that cross sections
and decay rates, which are defined in the physical region (i.e.\ the
region of time-like momenta), are calculable in QCD after a
``smearing'' or ``averaging'' procedure has been applied~\cite{PQW}.
In semileptonic decays, it is the integration over the lepton and
neutrino phase space that provides a smearing over the invariant
hadronic mass of the final state (so-called global duality). For
non-leptonic decays, on the other hand, the total hadronic mass is
fixed, and it is only the fact that one sums over many hadronic
states that provides an averaging (so-called local duality). Clearly,
local duality is a stronger assumption than global duality. It is
important to stress that quark-hadron duality cannot yet be derived
from first principles; still, it is a necessary assumption for many
applications of QCD. The validity of global duality has been tested
experimentally using data on hadronic $\tau$ decays~\cite{Maria}.
Some more formal attempts to address the problem of quark-hadron
duality can be found in Refs.~145,146.

Using the optical theorem, the inclusive decay width of a hadron
$H_b$ containing a $b$ quark can be written in the form
\begin{equation}\label{ImT}
   \Gamma(H_b\to X) = \frac{1}{m_{H_b}}\,\mbox{Im}\,
   \langle H_b|\,{\bf T}\,|H_b\rangle \,,
\end{equation}
where the transition operator ${\bf T}$ is given by
\begin{equation}
   {\bf T} = i\!\int{\rm d}^4x\,T\{\,
   {\cal L}_{\rm eff}(x),{\cal L}_{\rm eff}(0)\,\} \,.
\end{equation}
Inserting a complete set of states inside the time-ordered product,
we recover the standard expression
\begin{equation}
   \Gamma(H_b\to X) = {1\over 2 m_{H_b}}\,\sum_X\,
   (2\pi)^4\,\delta^4(p_H-p_X)\,|\langle X|\,{\cal L}_{\rm eff}\,
   |H_b\rangle|^2
\end{equation}
for the decay rate. For the case of semileptonic and non-leptonic
decays, ${\cal L}_{\rm eff}$ is the effective weak Lagrangian given
in (\ref{LFermi}), which in practice is corrected for short-distance
effects~\cite{AltM,Gail,cpcm3}$^-$\cite{cpcm5} arising from the
exchange of gluons with virtualities between $m_W$ and $m_b$. If some
quantum numbers of the final states $X$ are specified, the sum over
intermediate states is restricted appropriately. In the case of the
inclusive semileptonic decay rate, for instance, the sum would
include only those states $X$ containing a lepton-neutrino pair.

\begin{figure}[htb]
   \epsfxsize=7cm
   \centerline{\epsffile{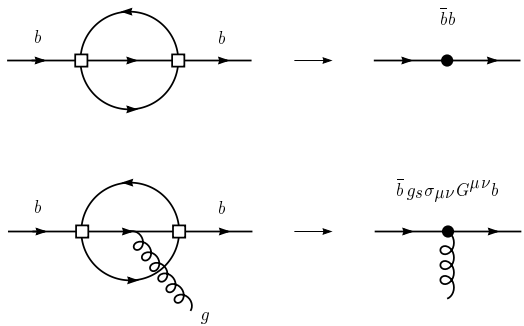}}
\caption{\label{fig:Toper}
Perturbative contributions to the transition operator ${\bf T}$
(left), and the corresponding operators in the OPE (right). The open
squares represent a four-fermion interaction of the effective
Lagrangian ${\cal L}_{\rm eff}$, while the black circles represent
local operators in the OPE.}
\end{figure}

In perturbation theory, some contributions to the transition operator
are given by the two-loop diagrams shown on the left-hand side in
Fig.~\ref{fig:Toper}. Because of the large mass of the $b$ quark, the
momenta flowing through the internal propagator lines are large. It
is thus possible to construct an OPE for the transition operator, in
which ${\bf T}$ is represented as a series of local operators
containing the heavy-quark fields. The operator with the lowest
dimension, $d=3$, is $\bar b b$. It arises by contracting the
internal lines of the first diagram. The only gauge-invariant
operator with dimension 4 is $\bar b\,i\rlap{\,/}D\,b$; however, the
equations of motion imply that between physical states this operator
can be replaced by $m_b\bar b b$. The first operator that is
different from $\bar b b$ has dimension 5 and contains the gluon
field. It is given by $\bar b\,g_s\sigma_{\mu\nu} G^{\mu\nu} b$. This
operator arises from diagrams in which a gluon is emitted from one of
the internal lines, such as the second diagram shown in
Fig.~\ref{fig:Toper}. For dimensional reasons, the matrix elements of
such higher-dimensional operators are suppressed by inverse powers of
the heavy-quark mass. Thus, any inclusive decay rate of a hadron
$H_b$ can be written as~\cite{Bigi}$^-$\cite{MaWe}
\begin{equation}\label{gener}
   \Gamma(H_b\to X_f) = {G_F^2 m_b^5\over 192\pi^3}\,
   \bigg\{ c_3^f\,\langle\bar b b\rangle_H
   + c_5^f\,{\langle\bar b\,g_s\sigma_{\mu\nu} G^{\mu\nu} b
   \rangle_H\over m_b^2} + \dots \bigg\} \,,
\end{equation}
where the prefactor arises naturally from the loop integrations,
$c_n^f$ are calculable coefficient functions (which also contain the
relevant CKM matrix elements) depending on the quantum numbers $f$ of
the final state, and $\langle O\rangle_H$ are the (normalized)
forward matrix elements of local operators, for which we use the
short-hand notation
\begin{equation}
   \langle O\rangle_H = {1\over 2 m_{H_b}}\,\langle H_b|\,
   O\,|H_b\rangle \,.
\end{equation}

In the next step, these matrix elements are systematically expanded
in powers of $1/m_b$, using the technology of the HQET. The result
is~\cite{FaNe,Bigi,MaWe}
\begin{eqnarray}
   \langle\bar b b\rangle_H &=& 1
    - {\mu_\pi^2(H_b)-\mu_G^2(H_b)\over 2 m_b^2} + O(1/m_b^3) \,,
    \nonumber\\
   \langle\bar b\,g_s\sigma_{\mu\nu} G^{\mu\nu} b\rangle_H
   &=& 2\mu_G^2(H_b) + O(1/m_b) \,,
\end{eqnarray}
where we have defined the HQET matrix elements
\begin{eqnarray}
   \mu_\pi^2(H_b) &=& {1\over 2 m_{H_b}}\,
    \langle H_b(v)|\,\bar b_v\,(i\vec D)^2\,b_v\,|H_b(v)\rangle \,,
    \nonumber\\
   \mu_G^2(H_b) &=& {1\over 2 m_{H_b}}\,
    \langle H_b(v)|\,\bar b_v {g_s\over 2}\sigma_{\mu\nu}
     G^{\mu\nu} b_v\,|H_b(v)\rangle \,.
\end{eqnarray}
Here $(i\vec D)^2=(i v\cdot D)^2-(i D)^2$; in the rest frame, this is
the square of the operator for the spatial momentum of the heavy
quark. Inserting these results into (\ref{gener}) yields
\begin{equation}\label{generic}
   \Gamma(H_b\to X_f) = {G_F^2 m_b^5\over 192\pi^3}\,
   \bigg\{ c_3^f\,\bigg( 1
   - {\mu_\pi^2(H_b)-\mu_G^2(H_b)\over 2 m_b^2} \bigg)
   + 2 c_5^f\,{\mu_G^2(H_b)\over m_b^2} + \dots \bigg\} \,.
\end{equation}
It is instructive to understand the appearance of the ``kinetic
energy'' contribution $\mu_\pi^2$, which is the gauge-covariant
extension of the square of the $b$-quark momentum inside the heavy
hadron. This contribution is the field-theory analogue of the Lorentz
factor $(1-\vec v_b^{\,2})^{1/2}\simeq 1-\vec k^{\,2}/2 m_b^2$, in
accordance with the fact that the lifetime, $\tau=1/\Gamma$, for a
moving particle increases due to time dilation.

The main result of the heavy-quark expansion for inclusive decay
rates is the observation that the free quark decay (i.e.\ the parton
model) provides the first term in a systematic $1/m_b$ expansion
\cite{Chay}. For dimensional reasons, the corresponding rate is
proportional to the fifth power of the $b$-quark mass. The
non-perturbative corrections, which arise from bound-state effects
inside the $B$ meson, are suppressed by at least two powers of the
heavy-quark mass, i.e.\ they are of relative order $(\Lambda_{\rm
QCD}/m_b)^2$. Note that the absence of first-order power corrections
is a consequence of the equations of motion, as there is no
independent gauge-invariant operator of dimension 4 that could appear
in the OPE. The fact that bound-state effects in inclusive decays are
strongly suppressed explains a posteriori the success of the parton
model in describing such processes \cite{ACCMM,Pasch}.

The hadronic matrix elements appearing in the heavy-quark expansion
(\ref{generic}) can be determined to some extent from the known
masses of heavy hadron states. For the $B$ meson, one finds that
\begin{eqnarray}\label{mupimuG}
   \mu_\pi^2(B) &=& - \lambda_1 = (0.3\pm 0.2)~\mbox{GeV}^2 \,,
    \nonumber\\
   \mu_G^2(B) &=& 3\lambda_2\approx 0.36~\mbox{GeV}^2 \,,
\end{eqnarray}
where $\lambda_1$ and $\lambda_2$ are the parameters appearing in the
mass formula (\ref{FNrela}). For the ground-state baryon $\Lambda_b$,
in which the light constituents have total spin zero, it follows that
\begin{equation}
   \mu_G^2(\Lambda_b) = 0 \,,
\end{equation}
while the matrix element $\mu_\pi^2(\Lambda_b)$ obeys the relation
\begin{equation}
   (m_{\Lambda_b}-m_{\Lambda_c}) - (\overline{m}_B-\overline{m}_D)
   = \Big[ \mu_\pi^2(B)-\mu_\pi^2(\Lambda_b) \Big]\,\bigg(
   {1\over 2 m_c} - {1\over 2 m_b} \bigg) + O(1/m_Q^2) \,,
\end{equation}
where $\overline{m}_B$ and $\overline{m}_D$ denote the spin-averaged
masses introduced in connection with (\ref{mbmc}). With the value of
$m_{\Lambda_b}$ given in (\ref{Lbmass}), this leads to
\begin{equation}\label{mupidif}
   \mu_\pi^2(B) - \mu_\pi^2(\Lambda_b) = (0.01\pm 0.03)~\mbox{GeV}^2
   \,.
\end{equation}
What remains to be calculated, then, is the coefficient functions
$c_n^f$ for a given inclusive decay channel. We shall now discuss
some of the most important applications of this general formalism.

\subsection{Determination of $|V_{cb}|$ from Inclusive Semileptonic
Decays}

The extraction of $|V_{cb}|$ from the inclusive semileptonic decay
rate of $B$ mesons is based on the general expression
(\ref{generic}), with the short-distance
coefficients~\cite{Bigi}$^-$\cite{MaWe}
\begin{eqnarray}
   c_3^{\rm SL} &=& |V_{cb}|^2 \Big[ 1 - 8 x^2 + 8 x^6 - x^8
    - 12 x^4\ln x^2 + O(\alpha_s) \Big] \,, \nonumber\\
   c_5^{\rm SL} &=& -6 |V_{cb}|^2 (1-x^2)^4 \,.
\end{eqnarray}
Here $x=m_c/m_b$, and $m_b$ and $m_c$ are the pole masses of the $b$
and $c$ quarks, defined to a given order in perturbation theory
\cite{Tarr}. The $O(\alpha_s)$ terms in $c_3^{\rm SL}$ are known
exactly~\cite{fgrefs}, while only partial calculations of
higher-order corrections exist~\cite{LSW,BaBB}. The theoretical
uncertainties in this determination of $|V_{cb}|$ are quite different
from those entering the analysis of exclusive decays. The main
sources are the dependence on the heavy-quark masses, unknown
higher-order perturbative corrections, and the assumption of global
quark-hadron duality. A conservative estimate of the total
theoretical error on the extracted value of $|V_{cb}|$
is~\cite{Beijing} $\delta|V_{cb}|/|V_{cb}|\approx 10\%$. Taking the
result of Ball et al.~\cite{BaBB} for the central value, and using
$\tau_B=(1.60\pm 0.03)\,$ps for the average $B$-meson
lifetime~\cite{Rich}, we find
\begin{equation}
   |V_{cb}| = (0.040\pm 0.004)\,\bigg(
   {\hbox{B}_{\rm SL}\over 10.8\%} \bigg)^{1/2}
   = (40\pm 1_{\rm exp}\pm 4_{\rm th}) \times 10^{-3} \,.
\end{equation}
In the last step, we have used $\hbox{B}_{\rm SL}=(10.8\pm 0.5)\%$
for the semileptonic branching ratio of $B$ mesons (see below). The
value of $|V_{cb}|$ extracted from the inclusive semileptonic width
is in excellent agreement with the value in (\ref{Vcbexc}) obtained
from the analysis of the exclusive decay $\bar B\to
D^*\ell\,\bar\nu$. This agreement is gratifying given the differences
of the methods used, and it provides an indirect test of global
quark-hadron duality. Combining the two measurements gives the final
result
\begin{equation}
   |V_{cb}| = 0.039\pm 0.002 \,.
\end{equation}
After $V_{ud}$ and $V_{us}$, this is the third-best known entry in
the CKM matrix.

\subsection{Semileptonic Branching Ratio for Decays into $\tau$
Leptons}

Semileptonic decays of $B$ mesons into $\tau$ leptons are of
particular importance, since they are sensitive probes of physics
beyond the Standard Model~\cite{Zoltau}. From the theoretical point
of view, the ratio of the semileptonic rates (or branching ratios)
into $\tau$ leptons and electrons can be calculated reliably. This
ratio is independent of the factor $m_b^5$, the hadronic parameter
$\lambda_1$, and CKM matrix elements. To order $1/m_b^2$, one
finds~\cite{incltau,Balk,Koyr}
\begin{equation}
   \frac{\hbox{B}(\bar B\to X\,\tau\,\bar\nu_\tau)}
        {\hbox{B}(\bar B\to X\,e\,\bar\nu_e)}
   = f(x_c,x_\tau) + \frac{\lambda_2}{m_b^2}\,g(x_c,x_\tau)
   = 0.22\pm 0.02 \,,
\label{Rtau}
\end{equation}
where $f$ and $g$ are calculable coefficient functions depending on
the mass ratios $x_c=m_c/m_b$ and $x_\tau=m_\tau/m_b$, as well as on
$\alpha_s(m_b)$. Two new measurements of the semileptonic branching
ratio of $b$ quarks, $\hbox{B}(b\to X\,\tau\,\bar\nu_\tau)$, have
been reported by the ALEPH and OPAL Collaborations at
LEP~\cite{Btau1,Btau2}. The weighted average is $(2.68\pm 0.28)\%$.
Normalizing this result to the LEP average value~\cite{Rich}
$\hbox{B}(b\to X\,e\,\bar\nu_e)=(10.95\pm 0.32)\%$, we obtain
\begin{equation}
   {\hbox{B}(b\to X\,\tau\,\bar\nu_\tau)\over
    \hbox{B}(b\to X\,e\,\bar\nu_\tau)} = 0.245\pm 0.027 \,,
\end{equation}
in good agreement with the theoretical prediction (\ref{Rtau}) for
$B$ mesons.

\subsection{Semileptonic Branching Ratio and Charm Counting}

The semileptonic branching ratio of $B$ mesons is defined as
\begin{equation}
   \hbox{B}_{\rm SL} = {\Gamma(\bar B\to X\,e\,\bar\nu)\over
   \sum_\ell \Gamma(\bar B\to X\,\ell\,\bar\nu) + \Gamma_{\rm had}
   + \Gamma_{\rm rare}} \,,
\end{equation}
where $\Gamma_{\rm had}$ and $\Gamma_{\rm rare}$ are the inclusive
rates for hadronic and rare decays, respectively. The main difficulty
in calculating $\hbox{B}_{\rm SL}$ is not in the semileptonic width,
but in the non-leptonic one. As mentioned previously, the calculation
of non-leptonic decay rates in the heavy-quark expansion relies on
the strong assumption of local quark-hadron duality.

Measurements of the semileptonic branching ratio have been performed
by various experimental groups, using both model-dependent and
model-inde\-pen\-dent analyses. The status of the results is
controversial, as there is a discrepancy between low-energy
measurements performed at the $\Upsilon(4s)$ resonance and
high-energy measurements performed at the $Z^0$ resonance. The
situation has been reviewed recently by Richman~\cite{Rich}, whose
numbers we shall use. The average value at low energies is
$\hbox{B}_{\rm SL}=(10.23\pm 0.39)\%$, whereas high-energy
measurements give $\hbox{B}_{\rm SL}(b)=(10.95\pm 0.32)\%$. The label
$(b)$ indicates that this value refers not to the $B$ meson, but to a
mixture of $b$ hadrons (approximately 40\% $B^-$, 40\% $B^0$, 12\%
$B_s$, and 8\% $\Lambda_b$). Assuming that the corresponding
semileptonic width $\Gamma_{\rm SL}(b)$ is close to that of $B$
mesons,\footnote{Theoretically, this is expected to be a very good
approximation.}
we can correct for this fact and find $\hbox{B}_{\rm
SL}=(\tau_B/\tau_b)\,\hbox{B}_{\rm SL}(b)=(11.23\pm 0.34)\%$, where
$\tau_b=(1.56\pm 0.03)\,$ps is the average lifetime corresponding to
the above mixture of $b$ hadrons. The discrepancy between the low-
and high-energy measurements of the semileptonic branching ratio is
therefore larger than three standard deviations. If we take the
average and inflate the error to account for this fact, we obtain
\begin{equation}\label{Bslval}
   \hbox{B}_{\rm SL} = (10.80\pm 0.51)\% \,.
\end{equation}
An important aspect in interpreting this result is charm counting,
i.e.\ the measurement of the average number $n_c$ of charm hadrons
produced per $B$ decay. Theoretically, this quantity is given by
\begin{equation}\label{ncdef}
   n_c = 1 + \hbox{B}(\bar B\to X_{c\bar c s'})
   - \hbox{B}(\bar B\to X_{{\rm no}\,c}) \,,
\end{equation}
where $\hbox{B}(\bar B\to X_{c\bar c s'})$ is the branching ratio for
decays into final states containing two charm quarks, and
$\hbox{B}(\bar B\to X_{{\rm no}\,c})\approx 0.02$ is the Standard
Model branching ratio for charmless
decays~\cite{Alta}$^-$\cite{Buch}. The average value obtained at low
energies is~\cite{Rich} $n_c=1.12\pm 0.05$, whereas high-energy
measurements give~\cite{ALEnc} $n_c=1.23\pm 0.07$. The weighted
average is
\begin{equation}
   n_c = 1.16\pm 0.04 \,.
\end{equation}

The naive parton model predicts that $\hbox{B}_{\rm SL}\approx 15\%$
and $n_c\approx 1.2$; however, it has been known for some time that
perturbative corrections could change these results
significantly~\cite{Alta}. With the establishment of the heavy-quark
expansion, the non-perturbative corrections to the parton model could
be computed, and their effect turned out to be very small. This led
Bigi et al.\ to conclude that values $\hbox{B}_{\rm SL}<12.5\%$
cannot be accommodated by theory~\cite{baff}. Later, Bagan et al.\
have completed the calculation of the $O(\alpha_s)$ corrections
including the effects of the charm-quark mass, finding that they
lower the value of $\hbox{B}_{\rm SL}$ significantly~\cite{BSLnew1}.
Their original analysis has recently been corrected in an erratum.
Here we shall present the results of an independent numerical
analysis using the same theoretical input~\cite{MNChris}. The
semileptonic branching ratio and $n_c$ depend on the quark-mass ratio
$m_c/m_b$ and on the ratio $\mu/m_b$, where $\mu$ is the scale used
to renormalize the coupling constant $\alpha_s(\mu)$ and the Wilson
coefficients appearing in the non-leptonic decay rate. The freedom in
choosing the scale $\mu$ reflects our ignorance of higher-order
corrections, which are neglected when the perturbative expansion is
truncated at order $\alpha_s$. We allow the pole masses of the heavy
quarks to vary in the range $m_b=(4.8\pm 0.2)$~GeV and $m_b-m_c=
(3.39\pm 0.06)$~GeV, corresponding to $0.25<m_c/m_b<0.33$. The value
of the difference $(m_b-m_c)$ is in accordance with (\ref{mQdif}).
Non-perturbative effects appearing at order $1/m_b^2$ in the
heavy-quark expansion are described by the single parameter
$\lambda_2$ in (\ref{lam2val}); the dependence on the parameter
$\lambda_1$ is the same for all inclusive decay rates and cancels out
in the predictions for $\hbox{B}_{\rm SL}$ and $n_c$. For the two
choices $\mu=m_b$ and $\mu=m_b/2$, we obtain~\cite{MNChris}
\begin{eqnarray}
   \hbox{B}_{\rm SL} &=& \cases{
    12.0\pm 1.0 \% ;& $\mu=m_b$, \cr
    10.9\pm 1.0 \% ;& $\mu=m_b/2$, \cr} \nonumber\\
   n_c &=& \cases{
    1.20\mp 0.06 ;& $\mu=m_b$, \cr
    1.21\mp 0.06 ;& $\mu=m_b/2$. \cr}
\end{eqnarray}
The uncertainties in the two quantities, which result from the
variation of $m_c/m_b$ in the range given above, are anticorrelated.
Notice that the semileptonic branching ratio has a stronger scale
dependence than $n_c$. By choosing a low renormalization scale,
values $\hbox{B}_{\rm SL}<12.5\%$ can easily be accommodated. This is
indeed not unnatural. Using the BLM scale-setting method~\cite{BLM},
it has been estimated that $\mu\gsim 0.32 m_b$ is an appropriate
scale to use in this case~\cite{LSW}.

\begin{figure}[htb]
   \epsfxsize=7cm
   \centerline{\epsffile{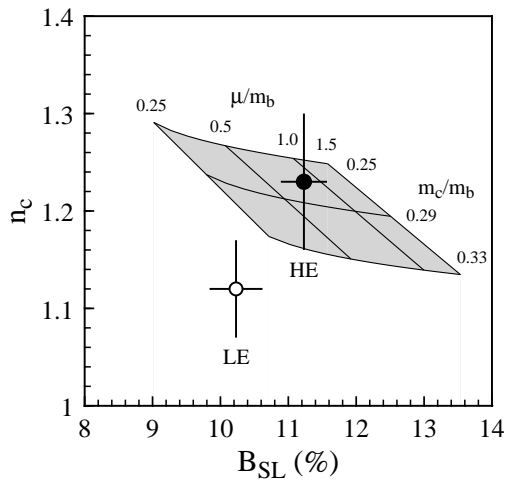}}
\caption{Theoretical prediction for the semileptonic branching ratio
and charm counting as a function of the quark-mass ratio $m_c/m_b$
and the renormalization scale $\mu$. The data points show the average
experimental values obtained in low-energy (LE) and high-energy (HE)
measurements, as discussed in the text.}
\label{fig:BSL}
\end{figure}

The combined theoretical predictions for the semileptonic branching
ratio and charm counting are shown in Fig.~\ref{fig:BSL}. They are
compared with the experimental results obtained from low- and
high-energy measurements. It has been argued that the combination of
a low semileptonic branching ratio and a low value of $n_c$ would
constitute a potential problem for the Standard Model~\cite{Buch}.
However, with the new experimental and theoretical numbers, only for
the low-energy measurements a discrepancy remains between theory and
experiment. Note that, with (\ref{ncdef}), our results for $n_c$ can
be used to calculate the branching ratio $\hbox{B}(\bar B\to X_{c\bar
c s'})$, which is accessible to a direct experimental determination.
Our prediction of $(22\pm 6)\%$ for this branching ratio agrees well
with the preliminary result reported by the CLEO Collaboration, which
is~\cite{Hons} $\hbox{B}(\bar B\to X_{c\bar c s'})=(23.9\pm 3.8)\%$.

Previous attempts to resolve the ``problem of the semileptonic
branching ratio'' have focused on four possibilities:
\begin{itemize}
\item
It has been argued that the experimental value of $n_c$ may depend on
model assumptions about the production of charm hadrons, which are
sometimes questionable~\cite{Buch,FDW}.
\item
It has been pointed out that the assumption of local quark-hadron
duality could fail in non-leptonic $B$
decays~\cite{Boyd,PaSt}$^-$\cite{Gui2}. If so, this will most likely
happen in the channel $b\to c\bar c s$, where the energy release,
$E=m_B - m_{X(c\bar cs)}$, is less than about 1.5~GeV. However, if
one assumes that sizeable duality violations occur only in this
channel, it is impossible to improve the agreement between theory and
experiment~\cite{Beijing}.
\item
Another possibility is that higher-order corrections in the $1/m_b$
expansion, which were previously thought to be negligible, give a
sizeable contribution. It has been argued that they could lower the
semileptonic branching ratio by up to 1\%, depending on the size of
some hadronic matrix elements~\cite{MNChris}.
\item
Finally, there is also the possibility to invoke New
Physics~\cite{Grza}$^-$\cite{CGG}. One may, for instance, consider
extensions of the Standard Model with enhanced flavour-changing
neutral currents, such as $b\to s\,g$. The effect of such a
contribution would be that both $\hbox{B}_{\rm SL}$ and $n_c$ are
reduced by a factor $(1+\eta \hbox{B}_{\rm SL}^{\rm SM})^{-1}$, where
$\eta=(\Gamma_{\rm rare}-\Gamma_{\rm rare}^{\rm SM})/ \Gamma_{\rm
SL}$. To obtain a sizeable decrease requires values $\eta\gsim 0.5$,
which are large (in the Standard Model, $\Gamma_{\rm rare}^{\rm SM}/
\Gamma_{\rm SL}\approx 0.2$), but not excluded by current
experiments.
\end{itemize}

\subsection{Lifetime Ratios of $b$ Hadrons}

The heavy-quark expansion shows that the lifetimes of all hadrons
containing a $b$ quark agree up to non-perturbative corrections
suppressed by at least two powers of $1/m_b$. In particular, it
predicts that
\begin{eqnarray}\label{taucrude}
   {\tau(B^-)\over\tau(B^0)} &=& 1 + O(1/m_b^3) \,,
    \nonumber\\
   {\tau(B_s)\over\tau(B_d)} &=& (1.00\pm 0.01) + O(1/m_b^3) \,,
    \nonumber\\
   {\tau(\Lambda_b)\over\tau(B^0)} &=& 1
    + {\mu_\pi^2(\Lambda_b)-\mu_\pi^2(B)\over 2 m_b^2}
    - c_G\,{\mu_G^2(B)\over m_b^2} + O(1/m_b^3) \nonumber\\
   &\approx& 0.98 + O(1/m_b^3) \,,
\end{eqnarray}
where~\cite{MNChris} $c_G\approx 1.1$, and we have used
(\ref{mupimuG}) and (\ref{mupidif}). The uncertainty in the value of
the ratio $\tau(B_s)/\tau(B_d)$ arises from unknown SU(3)-violating
effects in the matrix elements of $B_s$ mesons. The above theoretical
predictions may be compared with the average experimental values for
the lifetime ratios, which are~\cite{Rich}:
\begin{eqnarray}\label{taudata}
   {\tau(B^-)\over\tau(B^0)} &=& 1.06\pm 0.04 \,, \nonumber\\
   {\tau(B_s)\over\tau(B_d)} &=& 0.98 + 0.07 \,,
    \nonumber\\
   {\tau(\Lambda_b)\over\tau(B^0)} &=& 0.78\pm 0.04 \,.
\end{eqnarray}
Whereas the lifetime ratios of the different $B$ mesons are in good
agreement with the theoretical prediction, the low value of the
lifetime of the $\Lambda_b$ baryon is surprising.

\begin{figure}[htb]
   \epsfxsize=7cm
   \centerline{\epsffile{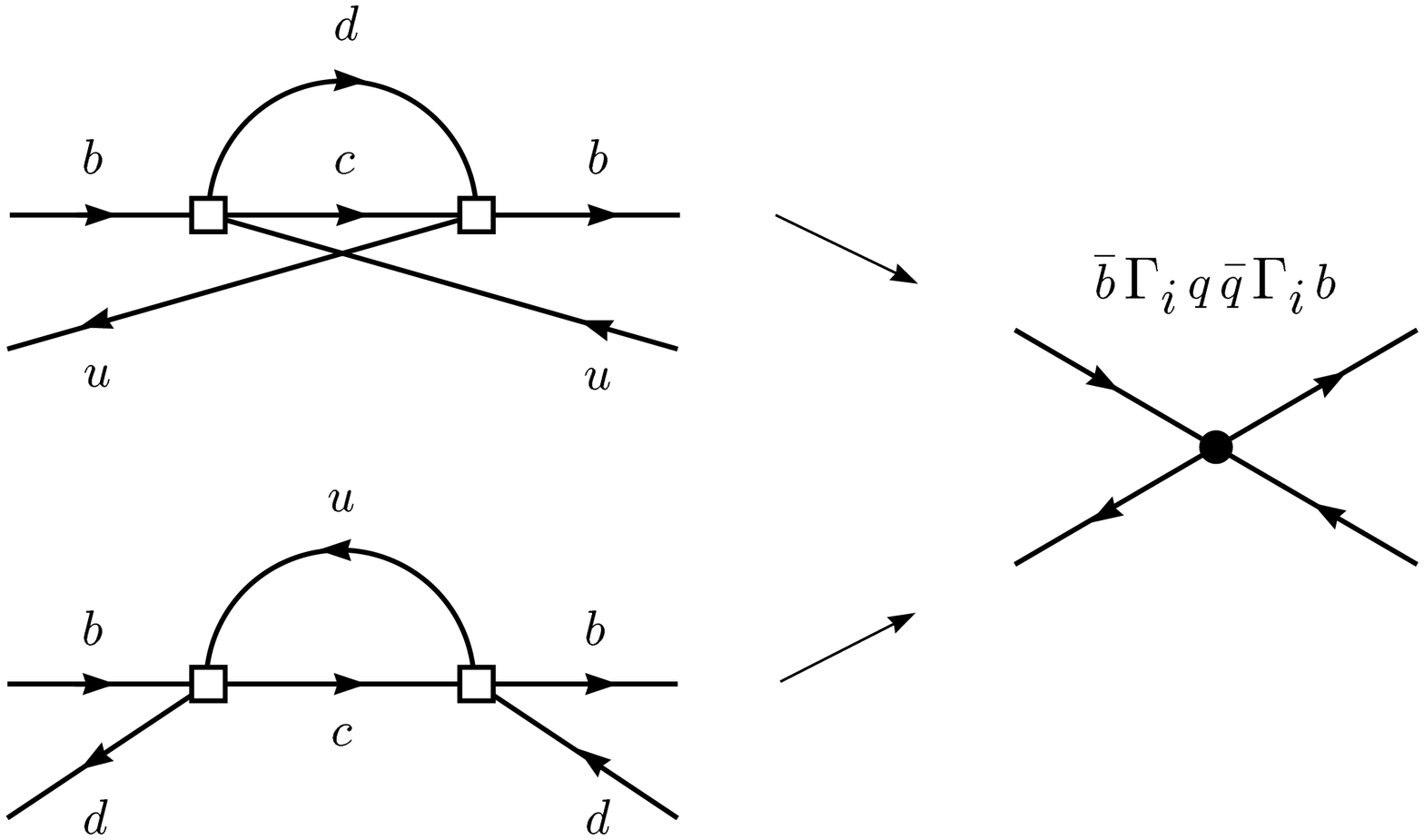}}
\caption{\label{fig:Tspec}
Spectator contributions to the transition operator ${\bf T}$ (left),
and the corresponding operators in the OPE (right). Here $\Gamma_i$
denotes some combination of Dirac and colour matrices.}
\end{figure}

To understand the structure of lifetime differences requires to go
further in the $1/m_b$ expansion~\cite{liferef}. Although at first
sight it appears that higher-order corrections could be safely
neglected given the smallness of the $1/m_b^2$ corrections, this
impression is erroneous for two reasons: first, at order $1/m_b^3$ in
the heavy-quark expansion for non-leptonic decay rates there appear
four-quark operators, whose matrix elements explicitly depend on the
flavour of the spectator quark(s) in the hadron $H_b$, and hence are
responsible for lifetime differences between hadrons with different
light constituents; secondly, these spectator effects receive a
phase-space enhancement factor of $O(16\pi^2)$ with respect to the
leading terms in the OPE~\cite{Beijing}. This can be seen from
Fig.~\ref{fig:Tspec}, which shows that the corresponding
contributions to the transition operator ${\bf T}$ arise from
one-loop rather than two-loop diagrams. The presence of this
phase-space enhancement factor leads to a peculiar structure of the
heavy-quark expansion for non-leptonic rates, which may be displayed
as follows:
\begin{eqnarray}
   \Gamma &=& \Gamma_0\,\Bigg\{ 1
    + x_2 \bigg( {\Lambda_{\rm QCD}\over m_b} \bigg)^2
    + x_3 \bigg( {\Lambda_{\rm QCD}\over m_b} \bigg)^3 + \dots
    \nonumber\\
   &&\mbox{}+ 16\pi^2\,\bigg[
    y_3 \bigg( {\Lambda_{\rm QCD}\over m_b} \bigg)^3
    + y_4 \bigg( {\Lambda_{\rm QCD}\over m_b} \bigg)^4 + \dots
    \bigg] \Bigg\} \,.
\end{eqnarray}
Here $\Gamma_0$ is the free quark decay rate, and $x_n$ and $y_n$ are
coefficients of order unity. Since it is conceivable that the terms
of order $16\pi^2\,(\Lambda_{\rm QCD}/m_b)^3$ could be larger than
the ones of order $(\Lambda_{\rm QCD}/m_b)^2$, it is important to
include them in the predictions for non-leptonic decay rates.
Moreover, there is a challenge to calculate the hadronic matrix
elements of the corresponding four-quark operators with high
accuracy.

In total, a set of four four-quark operators is induced by spectator
effects. They are
\begin{eqnarray}
   O_{\rm V-A}^q &=& \bar b\gamma_\mu (1-\gamma_5)q\,
    \bar q\gamma^\mu (1-\gamma_5)b \,, \nonumber\\
   O_{\rm S-P}^q &=& \bar b(1-\gamma_5)q\,\bar q(1+\gamma_5)b \,,
    \nonumber\\
   T_{\rm V-A}^q &=& \bar b\gamma_\mu (1-\gamma_5) t_a q\,
    \bar q\gamma^\mu (1-\gamma_5) t_a b \,, \nonumber\\
   T_{\rm S-P}^q &=& \bar b(1-\gamma_5) t_a q\,
    \bar q(1+\gamma_5) t_a b \,,
\end{eqnarray}
where $q$ is a light quark, and $t_a$ are the generators of colour
SU(3). In most previous analyses of spectator effects the hadronic
matrix elements of these operators have been estimated making
simplifying assumptions~\cite{liferef}$^-$\cite{ShiV}. For the matrix
elements between $B$-meson states the vacuum saturation
approximation~\cite{SVZ} has been assumed, i.e.\ the matrix elements
of the four-quark operators have been evaluated by inserting the
vacuum inside the current products. This leads to
\begin{eqnarray}\label{fact}
   \langle\bar B_q|\,O_{\rm V-A}^q\,|\bar B_q\rangle
   &=& \langle\bar B_q|\,O_{\rm S-P}^q\,|\bar B_q\rangle
    = f_{B_q}^2 m_{B_q}^2 \,, \nonumber\\
   \langle\bar B_q|\,T_{\rm V-A}^q\,|\bar B_q\rangle
   &=& \langle\bar B_q|\,T_{\rm S-P}^q\,|\bar B_q\rangle = 0 \,,
\end{eqnarray}
where $f_{B_q}$ is the decay constant of the $B_q$ meson, defined as
\begin{equation}
   \langle 0\,|\,\bar q\,\gamma^\mu\gamma_5\,b\,
   |\bar B_q(v)\rangle = i f_{B_q} m_{B_q} v^\mu \,.
\end{equation}
The vacuum saturation approximation has been criticized by
Chernyak~\cite{Chern}, who estimates that the corrections to it can
be as large as 50\%.

A model-independent analysis of spectator effects, which avoids
assumptions about hadronic matrix elements, can be performed if
instead of (\ref{fact}) one defines~\cite{MNChris}
\begin{eqnarray}\label{Biepsi}
   \langle\bar B_q|\,O_{\rm V-A}^q\,|\bar B_q\rangle
   &=& B_1\,f_{B_q}^2 m_{B_q}^2 \,, \nonumber\\
   \langle\bar B_q|\,O_{\rm S-P}^q\,|\bar B_q\rangle
   &=& B_2\,f_{B_q}^2 m_{B_q}^2 \,, \nonumber\\
   \langle\bar B_q|\,T_{\rm V-A}^q\,|\bar B_q\rangle
   &=& \varepsilon_1\,f_{B_q}^2 m_{B_q}^2 \,, \nonumber\\
   \langle\bar B_q|\,T_{\rm S-P}^q\,|\bar B_q\rangle
   &=& \varepsilon_2\,f_{B_q}^2 m_{B_q}^2 \,.
\end{eqnarray}
The values of the dimensionless hadronic parameters $B_i$ and
$\varepsilon_i$ are currently not known; ultimately, they may be
calculated using some field-theoretic approach such as lattice gauge
theory or QCD sum rules. The vacuum saturation approximation
corresponds to setting $B_i=1$ and $\varepsilon_i=0$ (at some scale
$\mu$, where the approximation is believed to be valid). For real
QCD, it is known that
\begin{equation}
   B_i = O(1) \,,\qquad \varepsilon_i=O(1/N_c) \,,
\end{equation}
where $N_c$ is the number of colours. Below, we shall treat $B_i$ and
$\varepsilon_i$ (renormalized at the scale $m_b$) as unknown
parameters. Similarly, the relevant hadronic matrix elements of the
four-quark operators between $\Lambda_b$-baryon states can be
parametrized by two parameters, $\widetilde B$ and $r$, where
$\widetilde B=1$ in the valence-quark approximation, in which the
colour of the quark fields in the operators is identified with the
colour of the quarks inside the baryon.

\subsubsection{Lifetime ratio for $B^-$ and $B^0$}

\begin{figure}[htb]
   \epsfxsize=9cm
   \centerline{\epsffile{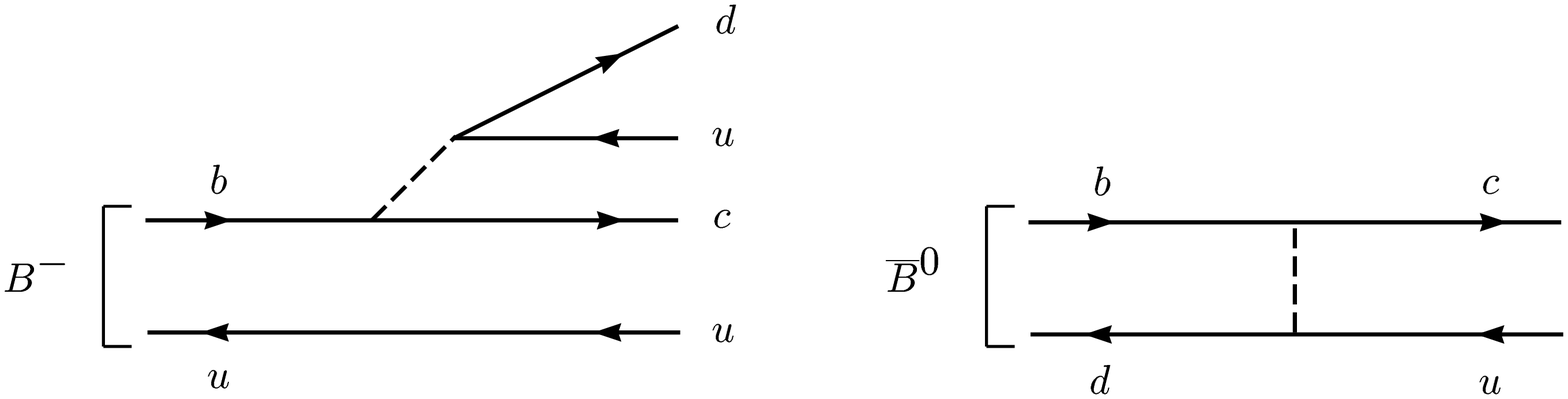}}
\caption{\label{fig:Wint}
Pauli interference and $W$ exchange contributions to the lifetimes of
the $B^-$ and the $\bar B^0$ mesons. The spectator effect in the
first diagram arises from the interference due to the presence of two
identical $\bar u$ quarks in the final state.}
\end{figure}

The lifetimes of the charged and neutral $B$ mesons differ because of
two types of spectator effects illustrated in Fig.~\ref{fig:Wint}.
They are referred to as Pauli interference and $W$
exchange~\cite{Gube}$^-$\cite{ShiV}. In the operator language, these
effects are represented by the hadronic matrix elements of the local
four-quark operators given in (\ref{Biepsi}). In fact, the diagrams
in Fig.~\ref{fig:Wint} can be obtained from those in
Fig.~\ref{fig:Tspec} by cutting the internal lines, which corresponds
to taking the imaginary part in (\ref{ImT}).

The explicit calculation of these spectator effects leads
to~\cite{MNChris}
\begin{equation}\label{DeltaGam}
   \Delta\Gamma_{\rm spec}(B_q) = {G_F^2 m_b^5\over 192\pi^3}\,
   |V_{cb}|^2\,16\pi^2\,{f_B^2\,m_B\over m_b^3}\,\zeta_{B_q} \,,
\end{equation}
where
\begin{equation}
   \zeta_{B^-}\approx -0.4 B_1 + 6.6\varepsilon_1 \,, \qquad
   \zeta_{B^0}\approx -2.2\varepsilon_1 + 2.4\varepsilon_2 \,.
\end{equation}
Note the factor of $16\pi^2$ in (\ref{DeltaGam}), which arises from
the phase-space enhancement of spectator effects. Given that the
parton-model result for the total decay width is
\begin{equation}
   \Gamma_{\rm tot}(B)\approx 3.7\times
   {G_F^2 m_b^5\over 192\pi^3}\,|V_{cb}|^2 \,,
\end{equation}
we see that the characteristic scale of the spectator contributions
is
\begin{equation}
   4\pi^2\,{f_B^2\,m_B\over m_b^3}\approx
   \bigg( {2\pi f_B\over m_b} \bigg)^2 \approx 5\% \,.
\end{equation}

The precise value of the lifetime ratio depends crucially on the size
of the hadronic matrix elements. Taking $f_B=200$~MeV for the decay
constant of the $B$ meson (see Ref.~24
and references
therein), i.e.\ absorbing the uncertainty in this parameter into the
definition of $B_i$ and $\varepsilon_i$, we find~\cite{MNChris}
\begin{eqnarray}
   {\tau(B^-)\over\tau(B^0)} &\approx& 1 + 0.03 B_1
    - 0.70\varepsilon_1 + 0.20\varepsilon_2 \nonumber\\
   &\approx& 1 + 0.05 \widehat{B}_1 - 0.75\widehat{\varepsilon}_1
    + 0.20\widehat{\varepsilon}_2 \,.
\end{eqnarray}
Here $B_i$ and $\varepsilon_i$ refer to a renormalization scale
$\mu=m_b$, whereas $\widehat{B}_i$ and $\widehat{\varepsilon}_i$
refer to a low renormalization point $\mu=\mu_{\rm had}$ chosen such
that $\alpha_s(\mu_{\rm had})=0.5$. The most striking feature of this
result is that the coefficients of the colour-octet operators $T_{\rm
V-A}$ and $T_{\rm S-P}$ are an order of magnitude larger than those
of the colour-singlet operator $O_{\rm V-A}$. As a consequence, the
vacuum insertion approximation, which was adopted in
Ref.~179
to predict that $\tau(B^-)/\tau(B^0)$ is larger
than unity by an amount of order 5\%, should not be trusted (not even
at a low renormalization point such as $\mu_{\rm had}$). With
$\varepsilon_i$ of order $1/N_c$, it is conceivable that the
non-factorizable contributions dominate the result. Thus, without a
detailed calculation of the parameters $\varepsilon_i$ no reliable
prediction can be obtained. Given our present ignorance about the
true values of the hadronic matrix elements, we must conclude that
even the sign of the sum of the spectator contributions cannot be
predicted. A lifetime ratio in the range
$0.8<\tau(B^-)/\tau(B^0)<1.2$ could be easily accommodated by theory.

In view of these considerations, the experimental fact that the
lifetime ratio turns out to be close to unity is somewhat of a
surprise. It implies a constraint on a certain combination of the
colour-octet matrix elements, which reads $\varepsilon_1 -
0.3\varepsilon_2=\mbox{few~\%}$.

\subsubsection{Lifetime ratio for $B_s$ and $B_d$}

The lifetimes of the two neutral mesons $B_s$ and $B_d$ differ
because spectator effects depend on the flavour of the light quark,
and moreover because the hadronic matrix elements in the two cases
differ by SU(3) symmetry-breaking corrections. It is difficult to
predict the sign of the net effect, but the magnitude cannot be
larger than one or two per cent~\cite{MNChris,liferef}. Hence
\begin{equation}
   {\tau(B_s)\over\tau(B_d)} = 1\pm O(1\%) \,,
\end{equation}
which is consistent with the experimental value in (\ref{taudata}).
Note that $\tau(B_s)$ denotes the average lifetime of the two $B_s$
states, whose individual lifetimes are expected to differ by a
sizeable amount~\cite{liferef,Martin}.

\subsubsection{Lifetime ratio for $\Lambda_b$ and $B^0$}

Although, as shown in (\ref{taucrude}), the lifetime differences
between heavy mesons and baryons start at order $1/m_b^2$ in the
heavy-quark expansion, the main effects are expected to appear at
order $1/m_b^3$. Therefore, for an estimate of the ratio
$\tau(\Lambda_b)/\tau(B^0)$ one needs the matrix elements of
four-quark operators between baryon states. Very little is known
about such matrix elements. Bigi et al.\ have adopted a simple
non-relativistic quark model to conclude that~\cite{liferef}
\begin{equation}
   {\tau(\Lambda_b)\over\tau(B^0)} = 0.90\mbox{--}0.95 \,.
\end{equation}
An even smaller lifetime difference has been obtained by
Rosner~\cite{Rosner}.

A model-independent analysis gives~\cite{MNChris}
\begin{equation}
   {\tau(\Lambda_b)\over\tau(\bar B^0)}\approx 0.98
   - 0.17\varepsilon_1 + 0.20\varepsilon_2
   - (0.012 + 0.021\widetilde B) r \,,
\end{equation}
where $\widetilde B$ and $r$ are expected to be positive and of order
unity. Given the structure of this result, it seems difficult to
explain the experimental value $\tau(\Lambda_b)/\tau(B^0) = 0.78\pm
0.04$ without violating the constraint $\varepsilon_1\approx
0.3\varepsilon_2$ mentioned above.\footnote{Another constraint arises
if one does not want to spoil the theoretical prediction for the
semileptonic branching ratio.}
Essentially the only possibility is to have $r$ of order 2--4 or so,
as there are good theoretical arguments why $\widetilde B$ cannot be
much larger than unity. On the other hand, in a constituent quark
picture, $r$ is the ratio of the wave functions determining the
probability to find a light quark at the location of the $b$ quark
inside the $\Lambda_b$ baryon and the $B$ meson, i.e.\
\begin{equation}
   r = {|\psi_{bq}^{\Lambda_b}(0)|^2
        \over |\psi_{b\bar q}^{B_q}(0)|^2} \,,
\end{equation}
and it is hard to see how this ratio could be much different from
unity.

In view of the above discussion, the observation of the short
$\Lambda_b$ lifetime remains a puzzle, whose explanation may lie
beyond the heavy-quark expansion. If the current experimental value
persists, one may have to question the validity of local quark-hadron
duality, which is assumed in the theoretical calculation of lifetimes
and non-leptonic inclusive decay rates~\cite{Gui2}.

\section{Concluding Remarks}

We have presented a review of the theory and phenomenology of
heavy-quark symmetry and its applications to the spectroscopy, weak
decays and lifetimes of hadrons containing a heavy quark. The
theoretical tools that allow us to perform quantitative calculations
are the heavy-quark effective theory and the $1/m_Q$ expansion. Our
hope is to have convinced the reader that heavy-flavour physics is a
rich and diverse area of research, which is at present characterized
by a fruitful interplay between theory and experiments. This has led
to many significant discoveries and developments. Heavy-quark physics
has the potential to determine many important parameters of the
electroweak theory and to test the Standard Model at low energies. At
the same time, it provides an ideal laboratory to study the nature of
non-perturbative phenomena in QCD.

The prospects for further significant developments in the field of
heavy-flavour physics look rather promising. With the approval of the
first asymmetric $B$ factories at SLAC and KEK, with ongoing
$B$-physics programs at the existing facilities at Cornell, Fermilab
and CERN, and with plans for future $B$ physics at HERA-B and the
LHC-B, there are $B$eautiful times ahead of us!

\end{document}